# Equilibrium of circular von-Kármán plate bonded with Kirchhoff rod*

Deepankar Das, Basant Lal Sharma

*Department of Mechanical Engineering, Indian Institute of Technology Kanpur, Kanpur, U. P. 208016, India, Phone:+91 512 259 6173, Fax:+91 512 259 7408*

## Abstract

A circular von-Kármán plate is considered bonded at its boundary to a circular Kirchhoff rod via a hinge-like junction. There is a mismatch of dimension between the rod and the plate boundary in their respective stress-free configurations. The process of gluing of the inextensible circular rod to the edge of the extensible plate causes the system to develop internal stress in the natural planar configuration. For some critical values of this mismatch the rod-plate system, as expected, shows buckling behaviour in the interior or at the boundary of plate. In this paper, the corresponding bifurcation problem is formulated using the mismatch parameter as a bifurcation parameter. The non-linear equations of equilibrium are linearized near the homogeneously deformed state and necessary conditions of bifurcation are solved for the critical geometric mismatch that is when non-trivial solutions of the linearized equations exist. These critical points are candidates to be classified as bifurcation points. Additionally, the semi-analytic nature of the analysis is exploited to perform a parametric study, relative to structure parameters in the problem formulation, of the critical points in the problem and the characteristic features are summarised. It is found that the rod plays an important role in determining the buckling behaviour when the plate's dimension is smaller than the looped rod, as in that case the rod buckles and the plate can bend or stretch depending on the planar or non-planar nature of buckled mode of the rod. When the plate is larger, the rod effectively stimulates a rigid Dirichlet boundary condition due to it's inextensibility; the critical points in this case are found to coincide with the case when buckling in the sole plate as a system is considered. Last but not the least, in the article, we also identify the symmetries corresponding to the null solutions at each critical point, which are exploited in a suitable post buckling analysis as part of a forthcoming manuscript.

## Introduction

Plates and shells, along with beams and rods, are ubiquitous in the realisation of modern technological structures (Carrera et al., 2011; Jones, 2006). The geometric aspect ratio, attributed to the 'thinness', or slenderness, plays a key role the theoretical analysis

*Manuscript based on the PhD thesis research carried out during 2018-2022 as part of the PhD dissertation written by the first author which is going to be submitted to the Indian Institute of Technology Kanpur.

*Email addresses:* `deepankd@iitk.ac.in` (Deepankar Das), `bls@iitk.ac.in` (Basant Lal Sharma )

of regimes of 'stability' (Euler, 1744; Lagrange, 1770; Timoshenko and Gere, 1961) of certain desirable configurations, such as straight beam, helical rod, cylindrical shell, or flat plate, in the presence of external conditions and fluctuating environment, that appear in such engineering applications (Bloom and Coffin, 2000). History of the developments behind physical understanding the deformation of such complex structures (Cosserat and Cosserat, 1909; Kármán, 1907; Donnell, 1934; Antman, 2013; Rubin and Cardon, 2002), its mathematical nature, and the key approaches to tackle problem of buckling (Budiansky, 1974; Koiter, 1981) are well known and found as part of several authoritative texts and popular expositions on this topic; see, for example, (Euler, 1744; Budiansky and Hutchinson, 2003). A glimpse of the modern challenges pertaining to the coupling between different physical fields in applications as well as in the emerging science of the mechanics of soft matter can be also found in the existing literature (Cao and Hutchinson, 2012; Vella et al., 2014; Durinck et al., 2023).

The mathematical analysis of particularly one dimensional rods and filaments, based on classical physical principles and the continuum models, can be found in a significant number of publications; for example, see (Antman and Kenney, 1981; Rubin and Cardon, 2002; Healey, 2002; Lestringant and Audoly, 2017). The problem on instabilities in such one dimensional structures has also been studied in the presence of other complexities arising from special interactions with environment or another structure (Greiner and Guggenberger, 1998; Paik and Thayamballi, 2000; Gaibotti et al., 2024).

The theoretical analysis certainly gets a way more complicated in the presence of several interacting layers and structures; for example, see the problems studied in composites and those associated with thermal effects (Emam and Nayfeh, 2009; Thornton, 1993), and many other interesting works along the same lines (Karam and Gibson, 1995; Freund, 2000; Tehrani et al., 2006; Coman, 2007; Michael and Kwok, 2007; Pi et al., 2007; Østergaard, 2008). Such a listing of references is far from exhaustive and has been mentioned taking into account a connection with the topic of the present article.

This paper investigates the equilibrium and loss of uniqueness of the equilibrium state for a system consisting of a circular von-Kármán plate with a Kirchhoff rod as the boundary. In addition to the different dimensionality of the two structures, i.e., the two-dimensional plate and the one-dimensional rod, there is a mismatch between the stress-free length of the plate boundary and that of the rod. This type of interaction between structures of different dimensionalities occurs in biological systems and soft robots, while the mismatch in lengths can be seen in systems involving growth and thermal stresses.

As an example of research problems combining structures of different dimensionalities is the analysis reported in (Karam and Gibson, 1995) as it deals with cylindrical cores with an elastic shell as the boundary, though we do not compare our results with theirs in any possible regime. A special mention of the remarkable problem of a soap film spanning a flexible filament (Giomi and Mahadevan, 2012; Biria and Fried, 2015) is, indeed, necessary. As another example of the research on a similar theme, that is between a two-dimensional and a one-dimensional structural element, is the one formulated in (Giomi and Mahadevan, 2012), which involves a soap film spanning an elastic loop. The stability analysis of equilibrium states of such systems was further investigated in (Biria and Fried, 2015) and (Maleki and Fried, 2013), with motivations rooted in modeling biological systems. Three-dimensional analogues of such a system can be observed in the structure of biological cells, as studied in works like (Sharma et al., 2023) and (Flormann et al., 2017).

Buckling problems of structures, related to the one we analyze in this article, have been studied in (Gaibotti et al., 2024) and (Rao and Rao, 2015), although they are limited to two-dimensional deformations and elastic support at the boundary rather than an elastic boundary. In this connection, the analysis of (Steigmann and Ogden, 1997) and (Steigmann and Ogden, 1999) also falls in this category as they investigated a system consisting of an elastic solid with a material boundary. From purely mathematical viewpoint, there are few examples of rigorous works which provide study of instabilities in not-so-straightforward geometric setting (Marthelot et al., 2017; Domokos et al., 1997). From applications viewpoint, in modern technology the small scales have started playing a major role. To this end the problem tackled in the present article is also anticipated to lead to a way to analyze stability problems when different class of structures are bonded together (Mousavi et al., 2013; Wang and Wang, 2013; Wang and Zhao, 2009; Rao and Rao, 2015).

In the back drop of structural and mathematical complexities present in several publications mentioned above, the goal of the present article is rather modest. From the viewpoint of understanding instabilities, which embodies solely the principles of solid mechanics in both one dimensional sub-structure and the two dimensional sub-structure to which it is bonded (for example, this statement can be read in contrast to the work of (Giomi and Mahadevan, 2012) and (Maleki and Fried, 2013) where only the one dimensional sub-structure is traditional solid like), a simple problem is considered. In this article, a circular von-Kármán elastic plate (Kármán, 1907; Donnell, 1934) is assumed to be bonded to a circular Kirchhoff elastic rod (Cosserat and Cosserat, 1909; Rubin and Cardon, 2002; Antman, 2013) at the plate boundary. A mismatch between the length of the rod and the perimeter of the plate, in their stress-free configurations, leads to the non-trivial problem of finding equilibrium configurations of the resulting elastic system. The aftermath of the process of glueing the circular rod to the edge of the plate is that the system develops internal stresses in its natural flat state; see Fig. 1 for an illustration.

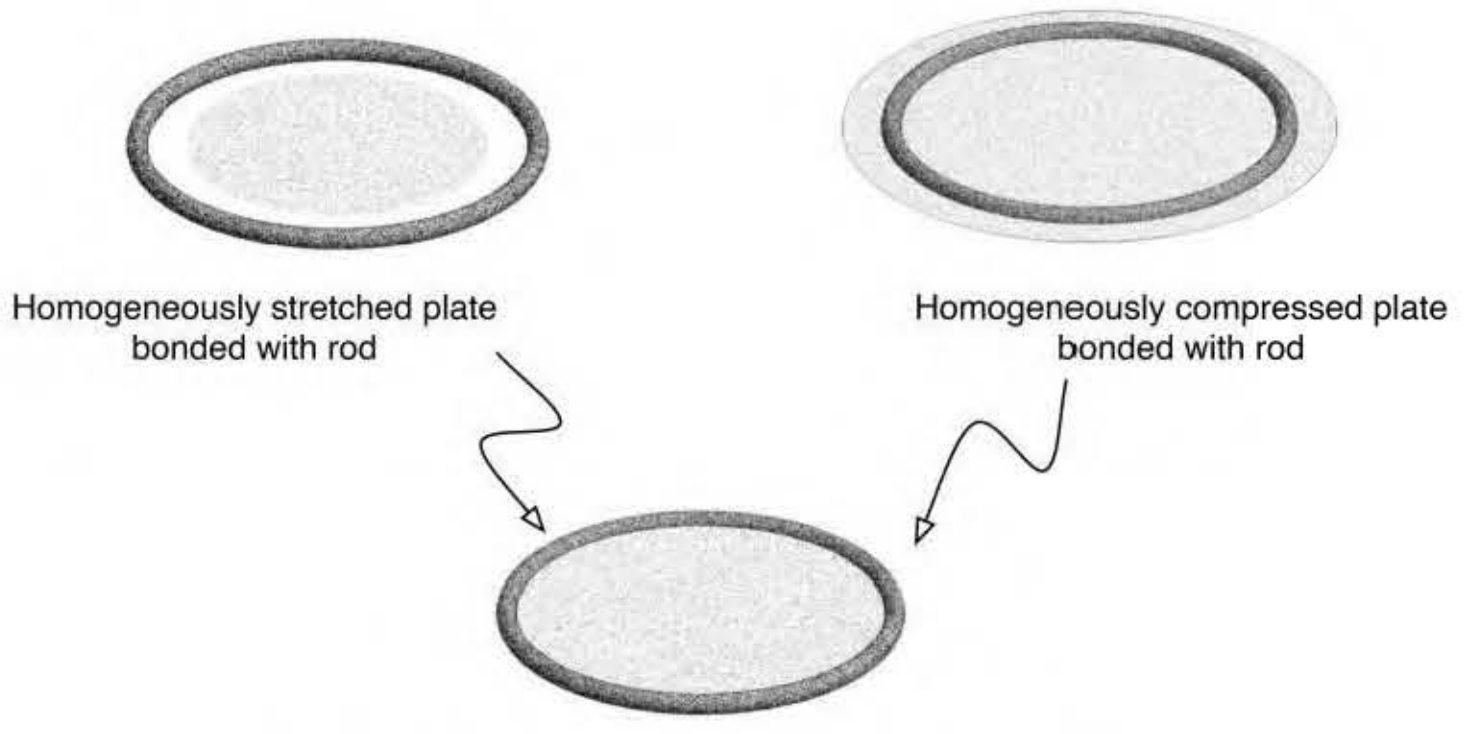

Figure 1: Bonding rod to the boundary of plate. The thickness of rod is exagerrated.

The mismatch of length between the stress-free plate and the boundary can arise from biological growth or from thermal expansion and contraction. Works like (Rodriguez et al., 1994) and (Ambrosi et al., 2011), which model stress-dependent growth in tissues and develop growth models for biological problems, could be employed in the context of length mismatch. Several works address thermal buckling and related phenomena: (Chiao and Lin, 2000) examines the thermal buckling of beams, while (Shariat and Eslami, 2006) and (Javaheri and Eslami, 2002) investigate the thermal buckling of graded plates, and (Thornton, 1993) provides a comprehensive overview of the thermal buckling of plates. Buckling behaviors due to thermal loading and growth motivate the selection of length mismatch as a bifurcation parameter in this paper.

Indeed, the natural flat state is the simplest configuration that corresponds to a homogeneously deformed plate when the materially homogeneous rod is homogeneous, inextensible and the plate is also materially homogeneous. Taking the ratio of the perimeter of stress-free plate to that of the length of the rod in its stress-free configurations as a mismatch parameter, for some critical value of this geometric mismatch, the simplest configuration of homogeneously deformed plate is expected to become unstable; see Fig. 2 for an illustration. In this paper, such a problem of finding the equilibria of the elastic system comprising plate bonded to rod is formulated as a non-linear operator over a suitable linear space (Healey, 1988; Kielhöfer, 2006). The non-linear operator, i.e. the system of non-linear equilibrium equations, is linearized and the null solutions of the linearized operator are found, that is so called buckling modes. This process actually satisfies the necessary conditions on the geometric mismatch of the system for which non-trivial buckling modes exist, and the criticial points are expected to be bifurcation points as they satisfy the necessary condition for the existence of a bifurcating branch (Kielhöfer, 2006). The semi-analytic nature of the analysis is exploited to perform a parametric study, relative to the structure parameters, of the problem and the key features of the corresponding analysis are mentioned. It is found that the rod plays an important role when the plate's dimension is smaller than the looped rod, in that case the rod buckles and the plate can bend or stretch depending on the buckled mode of the rod. When, plate is larger, the rod acts as a rigid Dirichlet boundary due to it's inextensibility, the critical points in this case are identical to only plate buckling. Furthermore, the symmetries corresponding to the null solutions are identified. These symmetries have been exploited on the lines of previous works (Healey, 1988) to perform a rigorous post-buckling analysis which is presently being written in the form of a separate exposition.

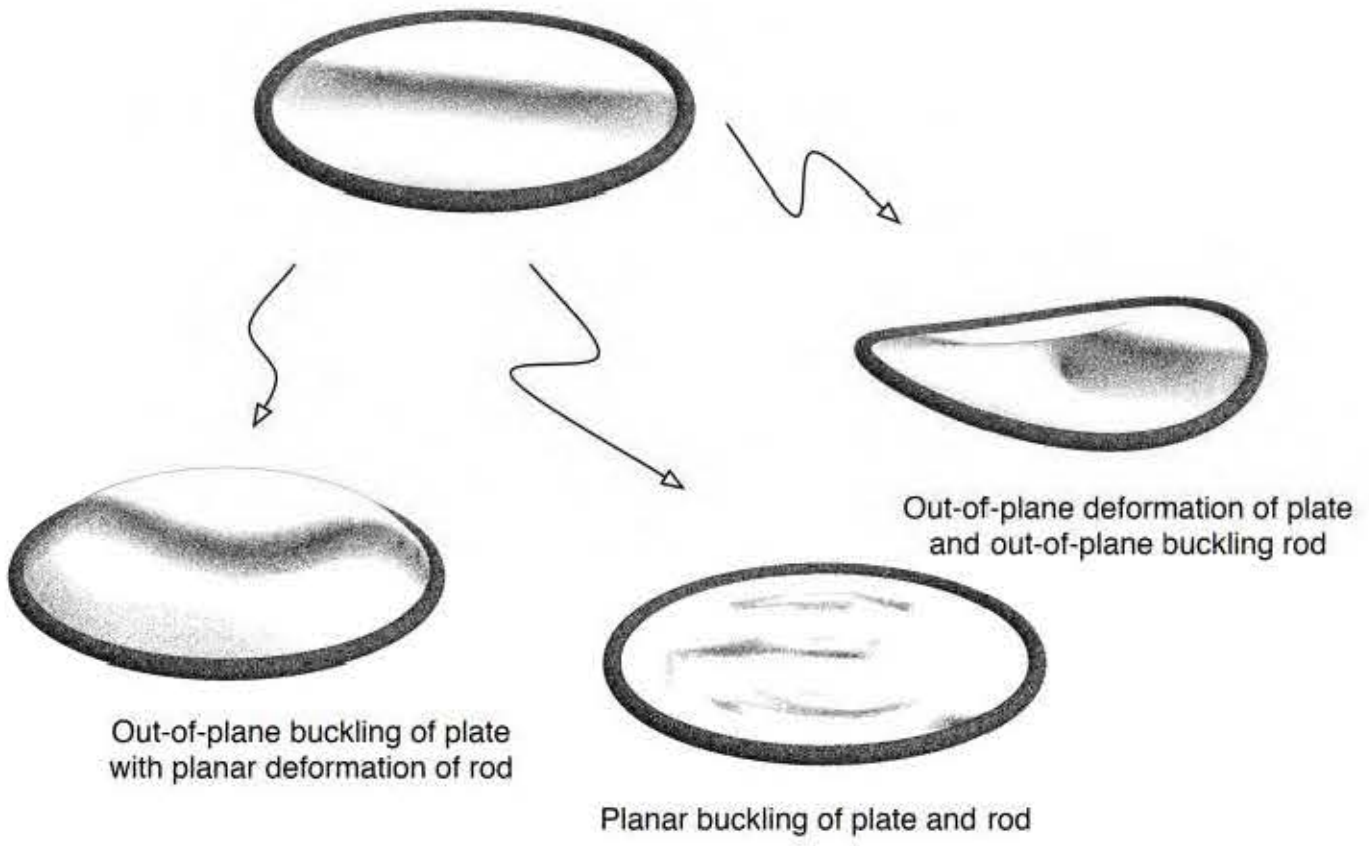

Figure 2: Schematic illustration of buckling instabilities in the system where a plate is bonded to a rod boundary. The thickness of rod is exagerrated.

From the viewpoint of the potential use of the results in this paper, the formulation of the problem, its analysis, and the semi-analytical results permit an application to the growth models of either plate or its boundary. This is a highly relevant area of research in currently available literature where several interesting models and theories are available, see, for example, (Nilsson et al., 1993; Rodriguez et al., 1994; Guillon et al., 2012; Moulton et al., 2013; Goriely, 2017; Lessinnes et al., 2017). The same comment applies to the case of thermally induced residual stress in the presence of geometric constraints (Shariat and Eslami, 2006; Cisternas et al., 2003; Chiao and Lin, 2000; Freund and Suresh, 2004;

Javaheri and Eslami, 2002). In addition, the spectral analysis of the associated linearized operator is closely connected to that of finding critical points that is for zero eigenvalue. There is an abundance of analytical and numerical results in slender structures related to this class of problems in the context of engineering applications; for example, see (Srivastava et al., 2003; Emam and Nayfeh, 2009). The model introduced in this article is also closely related to the existing well known models that incorporate surface and edge effects in continuous models (Gurtin and Ian Murdoch, 1975; Gurtin and Murdoch, 1978; Steigmann and Ogden, 1997, 1999). The thin film stability problems in presence of stiff substrate structures are also related in some sense (Freund and Suresh, 2004; Wang and Wang, 2013) as the rod-plate system permits the variation of relative softness of the two elements of the system. Applications of continuum models to biological structures and instability problems in soft matter is a rapidly emerging and highly relevant area of research (Ambrosi et al., 2011; Goriely, 2017; Almet et al., 2019; Sharma et al., 2023; Flormann et al., 2017). A specific range of choice of structure parameters allows potential usage of the system model assumed in the present article in such situations. On other hand, there are technological applications of the results from the viewpoint of energy harvesting and actuation; see, for example, (Cazottes et al., 2009; Pal et al., 2021) and (Qi et al., 2011; Cottone et al., 2012; Ansari and Karami, 2015; Hu and Burgueño, 2015). Lastly, it is worth mentioning that the topics such as the investigation of the nature of bifurcation points, detailed post-buckling analysis, and handling the issue of instabilities, is not explored in the present article. This is partly due to the limitations of keeping the exposition to basic and elementary form and, thus, deferring the more abstract analysis to a forthcoming article that emphasizes the last bit in a more circumspect manner.

*Outline*

The outline of the article is as follows. After fixing some important notational devices and symbols after this outline, in §1 we provide the necessary geometrical details for the system of plate bonded with rod and discuss the mechanics of the sole plate as well as that of the plate boundary as a special Cosserat rod, and eventually describe the hinge like bonding of the plate and the rod. Stored energy of the elastic system of plate bonded with rod is discussed in §2 where we give explicit expressions for the same from the point of a simple mathematical formulation. The equations of equilibrium of the plate bonded with rod are stated and briefly derived in §3. The necessary condition for bifurcation from the homogeneously deformed plate in order to bond with the special rod boundary with mismatched geometric parameter as a bifurcation parameter is used in §4 to obtain the in-plane as well as out-of-plane perturbed solution relative to the base solution; for this purpose an operator form of the equations of equilibrium is provided and the same is linearized relative to the base solution. A selection of special choice of material and geometric parameters is used in §5 to illustrate the semi-analytical and numerical results. A short discussion of main features in the article and few concluding remarks are summarized in §6. Lastly, there are four appendices in the article before the appearance of references, which deal with the first variation of the total energy, the unperturbed equilibrium equations as well as the perturbed equations and the complete listing of the various components of the linearized operator associated as domain equations and boundary conditions.

## Notation and mathematical preliminaries

The set of real numbers is denoted by $\mathbb{R}$ whereas the set of positive reals is denoted by $\mathbb{R}^+$. We also use the symbol $\mathbb{R}_{2\pi}$ to represent periodic interval of length $2\pi$. Let $O$ denote the origin in the three dimensional physical space (Euclidean point space) $\mathcal{E}$. Relative to $O$, the physical space $\mathcal{E}$ has the structure of three dimensional Euclidean (vector) space which is denoted by $\mathbb{R}^3$. Thus, the position vector of a point in $\mathcal{E}$ relative to $O$ is a vector in $\mathbb{R}^3$ and vice versa. We use boldface font to represent vectors (lower case) and tensors (upper case). $\mathbb{R}^3$ is equipped with the standard inner-product denoted by $\boldsymbol{a} \cdot \boldsymbol{b}$ for given $\boldsymbol{a}, \boldsymbol{b} \in \mathbb{R}^3$. $\mathbb{R}^3$ is also equipped with the standard cross-product denoted by $\boldsymbol{a} \wedge \boldsymbol{b}$ for given $\boldsymbol{a}, \boldsymbol{b} \in \mathbb{R}^3$ which satisfies the right-hand rule. Let $\{\boldsymbol{e}_1, \boldsymbol{e}_2, \boldsymbol{e}_3\}$ be a fixed (right handed) orthonormal basis, i.e. the standard basis, of $\mathbb{R}^3$. Thus, any vector $\boldsymbol{a} \in \mathbb{R}^3$ is uniquely expressed as $\boldsymbol{a} = a_i \boldsymbol{e}_i$ using the indicial summation convention. The components of vectors in $\mathbb{R}^3$ using the standard basis provide the natural coordinate system for the Euclidean point space $\mathcal{E}$ with respect to the origin $O$. The second order tensor $\boldsymbol{a} \otimes \boldsymbol{b}$, for given $\boldsymbol{a}, \boldsymbol{b} \in \mathbb{R}^3$, is defined as a linear transformation from $\mathbb{R}^3$ to $\mathbb{R}^3$ such that $(\boldsymbol{a} \otimes \boldsymbol{b})[\boldsymbol{c}] = \boldsymbol{a}(\boldsymbol{b} \cdot \boldsymbol{c})$ for all $\boldsymbol{c} \in \mathbb{R}^3$. The second order tensor $\mathtt{I}$ represents the identity tensor. The transpose of a second order tensor $\mathtt{A}$ is denoted by $\mathtt{A}^\top$ and is defined by the relation $\mathtt{A}\boldsymbol{v} \cdot \boldsymbol{v} = \boldsymbol{v} \cdot \mathtt{A}^\top \boldsymbol{v}$, for all $\boldsymbol{v}, \boldsymbol{v} \in \mathbb{R}^3$. The transpose of a second order tensor is defined in a way that $(\boldsymbol{a} \otimes \boldsymbol{b})^\top = \boldsymbol{b} \otimes \boldsymbol{a}$. A second order tensor $\mathtt{A}$ is called symmetric if it satisfies the property $\mathtt{A} = \mathtt{A}^\top$. The set $\{\boldsymbol{e}_i \otimes \boldsymbol{e}_j : i, j = 1, 2, 3\}$ forms the standard basis for second order tensors on $\mathbb{R}^3$. Thus, any second order tensor $\mathtt{A}$ is uniquely expressed as $\mathtt{A} = A_{ij} \boldsymbol{e}_i \otimes \boldsymbol{e}_j$ using the indicial summation convention. We also use the standard inner product for second order tensors which is denoted by ':' and defined as $\mathtt{A} : \mathtt{B} = A_{ij} B_{ij}$, for arbitrary second order tensors $\mathtt{A}$ and $\mathtt{B}$. Under this inner product, $\boldsymbol{a} \otimes \boldsymbol{b} : \boldsymbol{c} \otimes \boldsymbol{d} = (\boldsymbol{a} \cdot \boldsymbol{c})(\boldsymbol{b} \cdot \boldsymbol{d})$. Besides the vectors and tensors in three dimensions, we also employ analogous definitions for their restriction to the $\boldsymbol{e}_1$-$\boldsymbol{e}_2$ plane. The two dimensional Euclidean space as $\mathrm{span}(\boldsymbol{e}_1, \boldsymbol{e}_2)$ is denoted by $\mathbb{R}^2$, considered as a part of $\mathbb{R}^3$. The inner-product on $\mathbb{R}^2$ is also denoted by the same dot and defined as restriction of dot product on $\mathbb{R}^3$. Similar to the definition of the second order tensors on $\mathbb{R}^3$ (as linear transformations from $\mathbb{R}^3$ to $\mathbb{R}^3$), we also defined the second order tensors on $\mathbb{R}^2$ as linear transformations from $\mathbb{R}^2$ to $\mathbb{R}^2$. The transpose of such tensors is also defined in the same way. Thus, for example, a second order tensor $\mathbf{A}$ is called symmetric if it satisfies the property $\mathbf{A} = \mathbf{A}^\top$. The set of all symmetric second order tensors on $\mathbb{R}^2$ is denoted by Sym. The identity tensor on plane $\mathbb{R}^2$ is denoted by $\mathbf{I}$. The second order tensors are typically referred as tensors. In most of the article, we follow the standard notation in continuum mechanics and solid mechanics (Gurtin, 1982). The relevant definitions and notational choices appear in the article as and when required and their nature is clear from the context. The square brackets, in general, denote the linear action of an operator (say the gradient of a function, the first variation of functional, etc.) as clear from the context with an exception of the use of similar notation for a closed interval of the real line. For example, in case of second order tensors operating on vectors, this implies $\mathtt{A}\boldsymbol{a} = \mathtt{A}[\boldsymbol{a}] = A_{ij} a_j \boldsymbol{e}_j$. SO(3) denotes the group of all rotation tensors in three dimensions, i.e. given any $\mathtt{R} \in \mathrm{SO}(3)$, the defining relation is $\mathtt{R}\boldsymbol{a} \cdot \mathtt{R}\boldsymbol{b} = \mathtt{R}(\boldsymbol{a} \cdot \boldsymbol{b})$ and $\mathtt{R}\boldsymbol{a} \wedge \mathtt{R}\boldsymbol{b} = \boldsymbol{a} \wedge \boldsymbol{b}$ for all $\boldsymbol{a}, \boldsymbol{b} \in \mathbb{R}^3$. Given $\mathtt{R} \in \mathrm{SO}(3)$, it can be parametrized in terms of three scalar parameters. In this article, the rotation tensors SO(3) are parametrized by the Gibbs vector (Wilson and Gibbs, 1901). Several identities of vector and integral calculus are used in various parts of this article.

## 1. Geometry of plate bonded with rod

We start with the formulation of model system comprising a von-Kármán elastic plate bonded with a Kirchhoff elastic rod. In a way, the resulting elastic model structure is similar to the existing well known model (Gurtin and Ian Murdoch, 1975; Gurtin and Murdoch, 1978; Steigmann and Ogden, 1997, 1999), but there are important differences which result in a slightly different set of equations.

First we provide details for modelling the mechanics of a plate and the mechanics of a special Cosserat rod, and then provide the construction of an edge attached to the plate boundary (Kármán, 1907; Donnell, 1934; Jones, 2006; Janczewska and Zgorzelska, 2015; Bloom and Coffin, 2000) as a rod (Cosserat and Cosserat, 1909; Rubin and Cardon, 2002; Antman, 2013). To arrive at the model system, we use the homogeneously deformed plate and bond it to the rod. A bonding process is described that naturally invokes the definition of an intermediate configuration which is different from simply juxtaposing the stress-free configurations of the rod and plate together. This is because we have allowed a mismatch between the length of stress-free rod and the perimeter of stress-free plate.

### *1.1. Mechanics of plate*

A von-Kármán plate contained in the $\boldsymbol{e}_1$-$\boldsymbol{e}_2$ plane, i.e. $\mathbb{R}^2$, in its reference configuration, is considered. The stress-free reference configuration of the plate is (an open domain) represented by a solid circular disc $\widetilde{\Omega}$ of diameter $\mathfrak{P}$. The center of such a plate, in its reference configuration, coincides with the origin $O$ in the physical space $\mathcal{E}$. The deformation (placement) map of the plate is represented by

$$\widetilde{\boldsymbol{\chi}} : \widetilde{\Omega} \to \mathbb{R}^3. \tag{1.1}$$

The vector $\widetilde{\boldsymbol{y}} = \widetilde{\boldsymbol{\chi}}(\widetilde{\boldsymbol{x}}) \in \mathbb{R}^3$, for $\widetilde{\boldsymbol{x}} \in \widetilde{\Omega}$, is split in the form

$$\widetilde{\boldsymbol{y}} = \widetilde{\boldsymbol{x}} + \widetilde{\boldsymbol{v}}(\widetilde{\boldsymbol{x}}) + \widetilde{w}(\widetilde{\boldsymbol{x}})\boldsymbol{e}_3, \tag{1.2}$$

$$\text{where } \widetilde{\boldsymbol{v}} : \widetilde{\Omega} \to \mathbb{R}^2 \cong \text{span}(\boldsymbol{e}_1, \boldsymbol{e}_2) \ \text{ and } \ \widetilde{w} : \widetilde{\Omega} \to \mathbb{R}. \tag{1.3}$$

In physical terms, $\widetilde{\boldsymbol{v}}$ represents the in-plane displacement and $\widetilde{w}$ represents the out-of-plane displacement of the same material point in the von-Kármán plate. Corresponding to a deformation $\widetilde{\boldsymbol{\chi}}$ of the plate, the deformation gradient, $\widetilde{\mathtt{F}}$ $(:=\widetilde{\nabla}\widetilde{\boldsymbol{\chi}})$, is $\widetilde{\mathtt{F}} = \mathtt{I} + \widetilde{\nabla}\widetilde{\boldsymbol{v}} + \boldsymbol{e}_3 \otimes \widetilde{\nabla}\widetilde{w}$ on $\widetilde{\Omega}$. The exact expression of the Green's strain tensor for such $\widetilde{\mathtt{F}}$ is $\frac{1}{2}(\widetilde{\mathtt{F}}^\top\widetilde{\mathtt{F}} - \mathtt{I})$ which can be expanded as $\frac{1}{2}(\widetilde{\nabla}\widetilde{\boldsymbol{v}} + \widetilde{\nabla}\widetilde{\boldsymbol{v}}^\top + \widetilde{\nabla}\widetilde{\boldsymbol{v}}^\top\widetilde{\nabla}\widetilde{\boldsymbol{v}} + \widetilde{\nabla}\widetilde{w} \otimes \widetilde{\nabla}\widetilde{w})$. In these expressions, $\widetilde{\nabla}$ denotes the differential operator (gradient) with respect to the variable $\widetilde{\boldsymbol{x}}$ in the space $\mathbb{R}^2$ ($\equiv$ span$(\boldsymbol{e}_1, \boldsymbol{e}_2)$, as in (1.3)). In accordance with the von-Kármán model (Kármán, 1907), we consider

**Assumption 1.** *The second order terms in the gradient of in-plane deformation are ignored in the context of constitutive assumptions for the strain measures, while due to the 'small' out-of-plane displacement, only linear terms are considered for the curvature.*

In the von-Kármán model, the strain tensor

$$\widetilde{\mathbf{E}} : \widetilde{\Omega} \to \text{Sym} \cong \mathbb{R}^3 \tag{1.4}$$

is defined by

$$\widetilde{\mathbf{E}}:=\frac{1}{2}(\widetilde{\nabla}\widetilde{\boldsymbol{v}}+\widetilde{\nabla}\widetilde{\boldsymbol{v}}^{\top}+\widetilde{\nabla}\widetilde{w}\otimes\widetilde{\nabla}\widetilde{w}) \text{ on } \widetilde{\Omega}. \tag{1.5}$$

By Assumption 1, we employ the curvature (second order) tensor

$$\widetilde{\mathbf{K}}:\widetilde{\Omega}\to \mathrm{Sym}\cong\mathbb{R}^3,\quad \widetilde{\mathbf{K}}:=-\widetilde{\nabla}\widetilde{\nabla}\widetilde{w} \text{ on } \widetilde{\Omega}. \tag{1.6}$$

Further, as per the convention in plate theory, we consider

**Assumption 2.** *The von-Kármán plate is constructed from a three dimensional linear elastic, isotropic, homogeneous material body of thickness $h$, Young's modulus $E$, and Poisson's ratio $\nu$.*

The strain energy density in the von-Kármán plate is assumed to be a sum of the bending strain energy density $W_b$ (per unit reference area) and the membrane strain energy density $W_m$ (per unit reference area) (Healey et al., 2013). $W_m$ and $W_b$ are considered as functions of the in-plane strain tensor $\widetilde{\mathbf{E}}$ (1.5) and the curvature tensor $\widetilde{\mathbf{K}}$ (1.6) (both being considered relative to the stress-free configuration of plate), respectively, defined by the expressions

$$W_m(\widetilde{\mathbf{E}}):=6\,\mathfrak{C}\,W(\widetilde{\mathbf{E}}),\quad W_b(\widetilde{\mathbf{K}}):=\frac{1}{2}h^2\,\mathfrak{C}\,W(\widetilde{\mathbf{K}}), \tag{1.7}$$

where the common function $W$ is

$$W(\mathbf{A})=\nu(\mathrm{tr}\mathbf{A})^2+(1-\nu)\mathbf{A}:\mathbf{A}, \text{ for an arbitrary tensor } \mathbf{A},$$

$$\text{and}\quad \mathfrak{C}:=\frac{Eh}{12(1-\nu^2)}. \tag{1.8}$$

Note that $\mathfrak{C}>0$ and it has the units $[N]/[m]$. In (1.7), (1.8), in accordance with Assumption 2, $E>0$ is the Young's modulus (which has the units $[N]/[m]^2$) and $\nu\in(-1,1/2)$ is the Poisson's ratio while $h>0$ is the thickness (which has the units $[m]$) of the plate as a three dimensional plate-like body (Love, 2013; Kármán, 1907; Healey et al., 2013).

The discussion till this point concerns the interior region of the plate, while nothing has been specified about the behavior of the boundary (referred as the edge). In fact, the material behavior of the edge of the plate follows from a bonding process involving the perimeter of above mentioned plate and a circular Cosserat rod. The following details hold regarding the mechanical behaviour of the rod.

### *1.2. Mechanics of special Cosserat rod*

We consider a special Cosserat rod which is circular in its stress-free reference configuration. We suppose that the center-line of the stress-free reference configuration of the rod is represented by a circle $\widetilde{\Gamma}$ (as a subset of the three dimensional Euclidean point space $\mathcal{E}$) of diameter $\mathfrak{D}$ (note that $\mathfrak{D}$ can be different from the plate diameter $\mathfrak{P}$). Analogous to (1.1), let the deformation map of the rod center-line be denoted by

$$\widetilde{\boldsymbol{r}}:\widetilde{\Gamma}\to\mathbb{R}^3. \tag{1.9}$$

The material point $\widetilde{\boldsymbol{x}}\in\widetilde{\Gamma}$ corresponding to $\widetilde{s}\in\widetilde{\mathcal{I}}$ is written as $\widetilde{\boldsymbol{x}}(\widetilde{s})$, with $\widetilde{s}$ as the arclength parameter of the rod center-line $\widetilde{\Gamma}$.

Let $(r, \theta)$ represent the polar co-ordinates of a point in $\mathbb{R}^2$. Let

$$\boldsymbol{e}_r(r,\theta) \equiv \boldsymbol{e}_r(\theta) := \cos\theta \boldsymbol{e}_1 + \sin\theta \boldsymbol{e}_2, \quad \boldsymbol{e}_\theta(r,\theta) \equiv \boldsymbol{e}_\theta(\theta) := -\sin\theta \boldsymbol{e}_1 + \cos\theta \boldsymbol{e}_2, \tag{1.10}$$

denote the polar basis vectors in two dimensional space $\mathbb{R}^2$.

As we consider the rod to be circular in its stress-free reference configuration (centered at the origin $O$), for simplicity, the center-line in its reference configuration is described by

$$\widetilde{\boldsymbol{x}}(\widetilde{s}) = \frac{1}{2}\mathfrak{D}\ \boldsymbol{e}_r(\theta), \quad \widetilde{s} = \frac{1}{2}\mathfrak{D}\theta. \tag{1.11}$$

For convenience, we often identify $\widetilde{\boldsymbol{r}}(\widetilde{\boldsymbol{x}}(\widetilde{s}))$ as $\widetilde{\boldsymbol{r}}(\widetilde{s})$, and in a similar way several other fields defined on the rod.

The Cosserat triad of directors $\{\widetilde{\boldsymbol{d}_1}, \widetilde{\boldsymbol{d}_2}, \widetilde{\boldsymbol{d}_3}\}$ (forming a right handed orthonormal triad such that $\widetilde{\boldsymbol{d}_1} \wedge \widetilde{\boldsymbol{d}_2} \cdot \widetilde{\boldsymbol{d}_3} = +1$), attached to the center of the circular 'cross section' of the rod at each material point, are used to represent the orientation of cross section of the rod relative to the tangent of the center line. It is assumed that

$$\widetilde{\boldsymbol{d}_i} : \widetilde{\Gamma} \to \mathbb{R}^3, i = 1, 2, 3 \tag{1.12}$$

are smooth functions. A rotation tensor field

$$\widetilde{\mathtt{R}} : \widetilde{\Gamma} \to \mathrm{SO}(3) \tag{1.13}$$

is used to describe the orientation of $\{\widetilde{\boldsymbol{d}_1}, \widetilde{\boldsymbol{d}_2}, \widetilde{\boldsymbol{d}_3}\}$ relative to the reference directors $\{\widetilde{\boldsymbol{d}_1}^\dagger, \widetilde{\boldsymbol{d}_2}^\dagger, \widetilde{\boldsymbol{d}_3}^\dagger\}$, i.e.

$$\widetilde{\boldsymbol{d}_i} = \widetilde{\mathtt{R}}\widetilde{\boldsymbol{d}_i}^\dagger, \quad \text{for } i = 1, 2, 3, \tag{1.14}$$

at each material point on the rod. The function $\widetilde{Q} : \widetilde{\Gamma} \to \mathbb{R}^3 \times \mathrm{SO}(3)$ defined by $\widetilde{Q} := (\widetilde{\boldsymbol{r}}, \widetilde{\mathtt{R}})$, is used to describe the deformed configuration of the Coserrat rod. It is assumed that $\widetilde{Q}$ is a smooth function; the technical details are provided later.

Regarding the rod model, we consider

**Assumption 3.** *The rod is a Kirchhoff rod, i.e. unshearable and inextensible, composed of homogeneous, isotropic, and hyperelastic material.*

In view of above Assumption 3, the rod can undergo *only* bending and twisting deformation (Antman, 2013). The bending and twisting strain fields in the rod

$$\widetilde{\boldsymbol{\kappa}} : \widetilde{\Gamma} \to \mathbb{R}^3 \quad \text{and} \quad \widetilde{\tau} : \widetilde{\Gamma} \to \mathbb{R}, \tag{1.15}$$

respectively, are defined by

$$\widetilde{\boldsymbol{\kappa}} := \widetilde{\boldsymbol{d}_3}^\dagger \wedge (\widetilde{\mathtt{R}}^\top \widetilde{\mathtt{R}}' \widetilde{\boldsymbol{d}_3}^\dagger), \quad \widetilde{\tau} := \widetilde{\boldsymbol{d}_2}^\dagger \cdot (\widetilde{\mathtt{R}}^\top \widetilde{\mathtt{R}}' \widetilde{\boldsymbol{d}_1}^\dagger). \tag{1.16}$$

In (1.16), $()'$ refers to the derivative with respect to material arclength parameter $\widetilde{s}$ of the reference configuration $\widetilde{\Gamma}$.

As the rod is assumed to be circular in its stress-free reference configuration (centered at the origin $O$), for simplicity, the reference directors are chosen according to the following:

$$\widetilde{\boldsymbol{d}_1}^\dagger(\widetilde{s}) := \boldsymbol{e}_3, \quad \widetilde{\boldsymbol{d}_2}^\dagger(\widetilde{s}) := \boldsymbol{e}_r(\theta), \quad \widetilde{\boldsymbol{d}_3}^\dagger(\widetilde{s}) := \boldsymbol{e}_\theta(\theta), \tag{1.17}$$

where $\theta$ is given in terms $\widetilde{s}$ by $(1.11)_2$.

In accordance with above Assumption 3, the local form on $\widetilde{\Gamma}$ of the constraint of unshearability and inextensibility is expressed as

$$\widetilde{\boldsymbol{r}}' \cdot \widetilde{\mathrm{R}\boldsymbol{d}_3}^{\dagger} - 1 = 0, \tag{1.18}$$

$$\text{and} \quad \widetilde{\boldsymbol{r}}' \wedge (\widetilde{\mathrm{R}\boldsymbol{d}_3}^{\dagger}) = \boldsymbol{0}, \tag{1.19}$$

respectively, at each material point of the rod. Additionally, in order to obtain explicit expressions, the strain energy density of the rod (Antman, 2013) is assumed to be defined by

$$W_R(\widetilde{\boldsymbol{\kappa}}, \widetilde{\tau}) := \frac{1}{2}\left(\beta\widetilde{\boldsymbol{\kappa}} \cdot \widetilde{\boldsymbol{\kappa}} + \gamma\widetilde{\tau}^2\right), \tag{1.20}$$

where $\beta > 0$ is the bending modulus, $\gamma > 0$ denotes the twisting modulus (Antman, 2013), and the bending, twisting strains, that is $\widetilde{\boldsymbol{\kappa}}$, $\widetilde{\tau}$, are defined by (1.16).

### *1.3. Bonding of rod with the plate boundary*

In the process of bonding the above mentioned plate boundary and the circular rod, assuming $\mathfrak{P} \neq \mathfrak{D}$ in general, the plate is first homogeneously deformed (uniformly stretched if $\mathfrak{P} < \mathfrak{D}$ or compressed if $\mathfrak{P} > \mathfrak{D}$) so that its (circular) boundary coincides with the (circular) rod. We consider the situation corresponding to

**Assumption 4.** *The plate boundary and the rod are bonded together as a* frictionless hinge.

Since the rod is inextensible, during the gluing process the diameter of the rod doesn't change. The homogeneously deformed configuration of the plate is denoted by $\Omega$ (which is also a disk centered at $O$ and belonging to the same plane as $\widetilde{\Omega}$). As it is clear from the statements so far, the variables and fields defined on $\widetilde{\Omega}$ are decorated using $\widetilde{\cdot}$; however, after a few paragraphs, we make adjustments with respect to certain scalings and get rid of complicated adornments as $\widetilde{\cdot}$, a tactic resorted again and again in the manuscript as several classes of fields are involved while tackling the problem systematically and yet not testing levels of the reader's sanity.

Due to the inextensibility hypothesis (recall Assumption 3), the center line of the rod, after homogeneous deformation, concides with that in the reference configuration and since there is no twisting-bending deformation associated with this step of bonding $\widetilde{\boldsymbol{x}}(\widetilde{s}) \in \widetilde{\Gamma}$ coincides with $\boldsymbol{x}(s) \in \Gamma$ too while $\widetilde{\boldsymbol{d}}_i(\widetilde{s}) = \boldsymbol{d}_i(s)$. Recall that $\widetilde{s}$ is the arclength parameter of the reference rod (center line) $\widetilde{\Gamma}$ (same as $\Gamma$ due to inextensibility). As mentioned before, we identify $\widetilde{s}$ with $\widetilde{\boldsymbol{x}} \in \widetilde{\Gamma}$ and simply write, to avoid clutter in notation, $\widetilde{\boldsymbol{d}}_i(\widetilde{s})$ in place of $\widetilde{\boldsymbol{d}}_i(\widetilde{\boldsymbol{x}}(\widetilde{s}))$.

Post above mentioned homogenous deformation of the plate, for each $\widetilde{\boldsymbol{x}} \in \widetilde{\Omega}$, the corresponding location in the 'intermediate' configuration $\boldsymbol{x} \in \Omega$ is given by

$$\boldsymbol{x} = \frac{\widetilde{\boldsymbol{x}}}{\lambda}, \quad \text{where } \lambda := \frac{\mathfrak{P}}{\mathfrak{D}}. \tag{1.21}$$

Note that the parameter $\lambda$, defined above, takes only positive values, i.e.

$$\lambda \in \mathbb{R}^+. \tag{1.22}$$

In the foregoing analysis $\lambda$ plays a key role and it is treated as a bifurcation parameter while seeking a family of equilibria of the system comprised of plate bonded to rod.

Let

$$\boldsymbol{v} : \Omega \to \text{span}(\boldsymbol{e}_1, \boldsymbol{e}_2) \text{ and } w : \Omega \to \mathbb{R} \tag{1.23}$$

represent the in-plane displacement field and the out-of-plane displacement field defined relative to the intermediate configuration $\Omega$. This is similar to the definition (1.3). In fact, the displacement fields defined on $\Omega$ and $\widetilde{\Omega}$ are related by (using (1.21))

$$\boldsymbol{v}(\boldsymbol{x}) = \widetilde{\boldsymbol{v}}(\widetilde{\boldsymbol{x}}) - (\frac{1}{\lambda} - 1)\widetilde{\boldsymbol{x}}, \quad w(\boldsymbol{x}) = \widetilde{w}(\widetilde{\boldsymbol{x}}). \tag{1.24}$$

Due to (1.22), note that $\lambda$ is non-zero. The boundary of the homogeneously deformed plate $\Omega$ and the center-line of the rod $\Gamma$ (same as $\widetilde{\Gamma}$ so that $s$ can be also used in place of $\widetilde{s}$) coincide. In other words,

$$\Gamma = \partial\Omega. \tag{1.25}$$

The continuity of the displacement field of the rod and the plate, at the plate boundary, is specified by the condition

$$\widetilde{\boldsymbol{r}}(\widetilde{s}) - \boldsymbol{x}(s) = \boldsymbol{v}(\boldsymbol{x}(s)) + w(\boldsymbol{x}(s))\boldsymbol{e}_3, \text{ for all } s \in \mathcal{I}. \tag{1.26}$$

Note that

$$\boldsymbol{x}(s) = \widetilde{\boldsymbol{x}}(\widetilde{s}) \in \Gamma \text{ and } s = \widetilde{s} \in \mathcal{I}. \tag{1.27}$$

## 2. Stored energy of the elastic system of plate bonded with rod

It is assumed that the rod and plate are bonded in such a way that the rod-plate junction does not resist the plate bending in any direction normal to the center-line of the rod (that is Assumption 4). That is the interface between the rod and the plate has a (frictionless) hinge-like coupling. The details of such interface condition are naturally obtained via the variational framework to obtain the equations of equilibrium as described in the following.

The equations of equilibrium of the elastic system of plate bonded with rod are obtained using an application of the first variation based conditions of equilibrium using the total potential energy. Some details of the first variation of the total stored energy appear in Appendix A.

In the problem studied in this article, the total potential energy is equal to the stored energy due to the absence of external forces and moments on the elastic system.

The total stored energy of the mechanical system composed of the rod and plate is a sum of the elastic energy stored in the plate and the rod using the definitions of the energy densities (1.7), (1.20). The stored elastic energy of the plate and the rod, respectively, is defined by the energy functionals

$$\mathcal{V}_P(\widetilde{\boldsymbol{v}}, \widetilde{w}) := \int_{\widetilde{\Omega}} \left( W_m(\widetilde{\mathbf{E}}) + W_b(\widetilde{\mathbf{K}}) \right) \mathrm{d}\widetilde{A}, \quad \mathcal{V}_R(\widetilde{\mathrm{R}}) := \int_{\widetilde{\Gamma}} W_R(\widetilde{\boldsymbol{\kappa}}, \widetilde{\tau}) \mathrm{d}\widetilde{l}. \tag{2.1}$$

For specific calculations, for example, it can be assumed that $\mathrm{d}\widetilde{A} \simeq \mathrm{d}\widetilde{x}_1 \mathrm{d}\widetilde{x}_2$ and $\mathrm{d}\widetilde{l} \simeq \mathrm{d}\widetilde{s}$ in (2.1).

**Remark 1.** *For brevity the functional details are not mentioned above in (2.1), but it is implicit that $\mathcal{V}_P$ is a functional of the displacement field (1.2) of the plate and $\mathcal{V}_R$ depends on the rotation part of the deformation field $\widetilde{Q}:=(\widetilde{\boldsymbol{r}}, \widetilde{\mathtt{R}})$ of the rod (recall the statement following (1.14)). Note that the Lagrangian function for the functional $\mathcal{V}_P$ depends only the first derivative of the in-plane displacement, while it involves the first derivative* and *the second derivative of the out-of-plane displacement via (1.5) and (1.6). The Lagrangian function for the functional $\mathcal{V}_R$ depends on the rotation tensor field and its first derivative via (1.16).*

In order to account for the inextensibility constraint (1.18) and the unshearability constraint (1.19) of the rod, respectively, the Lagrange multipliers

$$n_{||} : \widetilde{\Gamma} \to \mathbb{R} \quad \text{and} \quad \boldsymbol{n}^L : \widetilde{\Gamma} \to \mathbb{R}^3 \tag{2.2}$$

are introduced. Since the constraint (1.19) has non-zero components only along $\boldsymbol{d}_1$ and $\boldsymbol{d}_2$, the corresponding Lagrange multiplier $\boldsymbol{n}^L$ also belongs to the $\boldsymbol{d}_1$-$\boldsymbol{d}_2$ plane. Therefore, an additional restriction is that

$$\boldsymbol{n}^L \cdot (\widetilde{\mathtt{R}\boldsymbol{d}_3}^{\dagger}) = 0, \text{ and, therefore, } \Upsilon : \widetilde{\Gamma} \to \mathbb{R}, \tag{2.3}$$

is introduced as an additional Lagrange multiplier.

**Remark 2.** *Equivalent to (1.19), we have the pair of constraints $\widetilde{\boldsymbol{r}}' \cdot \widetilde{\mathtt{R}\boldsymbol{d}_1} = 0, \widetilde{\boldsymbol{r}}' \cdot \widetilde{\mathtt{R}\boldsymbol{d}_2} = 0$, which utilizes only two scalar Lagrange multipliers. The resulting analysis remains the same though the details are not pursued, in this article, with this less awkward choice.*

The (augmented) potential energy functional (including the contribution of the imposed constraints) of the entire mechanical system is defined by

$$\mathcal{V} := \mathcal{V}_P + \mathcal{F}_R, \tag{2.4}$$

using $(2.1)_1$, while $\mathcal{F}_R$ is a sum of the stored energy $\mathcal{V}_R$ $(2.1)_2$ of the rod and certain terms describing constraints, namely the inextensibility constraint (1.18), the unshearability constraint (1.19) and the constraint $(2.3)_1$; in particular,

$$\text{where } \mathfrak{C}^{-1}\mathcal{V}_P := \int_{\widetilde{\Omega}} 6\left(\nu(\operatorname{tr}\widetilde{\mathbf{E}})^2 + (1-\nu)\widetilde{\mathbf{E}} : \widetilde{\mathbf{E}}\right) \mathrm{d}\widetilde{A} + + \int_{\widetilde{\Omega}} \frac{h^2}{2}\left(\nu(\operatorname{tr}\widetilde{\mathbf{K}})^2 + (1-\nu)\widetilde{\mathbf{K}} : \widetilde{\mathbf{K}}\right) \mathrm{d}\widetilde{A}, \tag{2.5}$$

$$\mathcal{F}_R := \int_{\widetilde{\Gamma}} \frac{1}{2}\left(\beta\widetilde{\boldsymbol{\kappa}} \cdot \widetilde{\boldsymbol{\kappa}} + \gamma\widetilde{\tau}^2\right) \mathrm{d}\widetilde{l} + \int_{\widetilde{\Gamma}} n_{||}(\widetilde{\boldsymbol{r}}' \cdot \widetilde{\mathtt{R}\boldsymbol{d}_3}^{\dagger} - 1)\mathrm{d}\widetilde{l} + + \int_{\widetilde{\Gamma}} \boldsymbol{n}^L \cdot (\widetilde{\boldsymbol{r}}' \wedge \left(\widetilde{\mathtt{R}\boldsymbol{d}_3}^{\dagger}\right))\mathrm{d}\widetilde{l} + \int_{\widetilde{\Gamma}} \Upsilon \boldsymbol{n}^L \cdot (\widetilde{\mathtt{R}\boldsymbol{d}_3}^{\dagger})\mathrm{d}\widetilde{l}. \tag{2.6}$$

Note that $\beta$ (resp. $\gamma$) has the units $[N][m]^2$. Also $\boldsymbol{n}^L$ (resp. $n_{||}$) has the units $[N]$.

Using the expanded form, to emphasize the variables and fields involved, it is clear that the first term is $\mathcal{V}_P(\boldsymbol{v}, w)$ and the second term is $\mathcal{F}_R \equiv \mathcal{F}_R(\boldsymbol{v}, w, \mathtt{R}, n_{||}, \boldsymbol{n}^L, \Upsilon)$ in (2.6), so that, in (2.4), $\mathcal{V} \equiv \mathcal{V}(\boldsymbol{v}, w, \mathtt{R}, n_{||}, \boldsymbol{n}^L, \Upsilon)$, where $n_{||}, \boldsymbol{n}^L, \Upsilon$ are the Lagrange multipliers mentioned in (2.2), $(2.3)_2$.

## 3. Equations of equilibrium of plate bonded with rod

At equilibrium, the first variation of the (augmented) potential energy functional (2.4) is zero for all admissible variations $\delta\boldsymbol{v}$, $\delta w$, $\delta\boldsymbol{\theta}$, $\delta n_{||}$, $\delta\boldsymbol{n}_{\perp}$, and $\delta\Upsilon$. The Euler-Lagrange equations (Gelfand et al., 2000; Giaquinta and Hildebrandt, 2004) are found to be (see Appendix A for details)

$$\nabla\cdot\mathbf{N} = \mathbf{0}, \tag{3.1a}$$

$$\nabla\cdot(\frac{1}{\lambda}\mathbf{N}\nabla w + \nabla\cdot\mathbf{M}) = 0, \tag{3.1b}$$

on $\Omega$, i.e. in the interior of the plate, where $\mathbf{N}:\Omega\to\mathrm{Sym}\cong\mathbb{R}^3$ and $\mathbf{M}:\Omega\to\mathrm{Sym}\cong\mathbb{R}^3$ are defined by

$$\mathbf{N}(\boldsymbol{x}) := \lambda\widetilde{\mathbf{N}}(\widetilde{\boldsymbol{x}}), \quad \mathbf{M}(\boldsymbol{x}) := \widetilde{\mathbf{M}}(\widetilde{\boldsymbol{x}}), \tag{3.2}$$

with $\widetilde{\boldsymbol{x}}$ and $\boldsymbol{x}$ related by (1.21) and

$$\widetilde{\mathbf{N}} := 6\mathfrak{C}\Big(\nu(2\widetilde{\nabla}\cdot\widetilde{\boldsymbol{v}} + \widetilde{\nabla}\widetilde{w}\cdot\widetilde{\nabla}\widetilde{w})\mathbf{I} + +(1-\nu)\left(\widetilde{\nabla}\widetilde{\boldsymbol{v}} + \widetilde{\nabla}\widetilde{\boldsymbol{v}}^{\top} + \widetilde{\nabla}\widetilde{w}\otimes\widetilde{\nabla}\widetilde{w}\right)\Big), \tag{3.3a}$$

$$\widetilde{\mathbf{M}} := -h^2\mathfrak{C}\left(\nu\widetilde{\Delta}\widetilde{w}\mathbf{I} + (1-\nu)\widetilde{\nabla}^2\widetilde{w}\right), \tag{3.3b}$$

while on $\Gamma=\partial\Omega$, i.e. on the boundary of $\Omega$ (the rod) (1.25), we have

$$\boldsymbol{m}' + (\boldsymbol{n}^L\wedge\boldsymbol{d}_3)\wedge\boldsymbol{r}' - \Upsilon\boldsymbol{d}_3\wedge\boldsymbol{n}^L = \mathbf{0}, \tag{3.4a}$$

$$(\mathtt{I} - \boldsymbol{e}_3\otimes\boldsymbol{e}_3)\boldsymbol{n}' - \mathbf{N}\boldsymbol{l} = \mathbf{0}, \tag{3.4b}$$

$$\boldsymbol{n}'\cdot\boldsymbol{e}_3 - ((\frac{1}{\lambda}\mathbf{N}\nabla w + \nabla\cdot\mathbf{M})\cdot\boldsymbol{l} + \nabla(\boldsymbol{t}\cdot\mathbf{M}\boldsymbol{l})\cdot\boldsymbol{t}) = 0, \tag{3.4c}$$

$$\boldsymbol{l}\cdot\mathbf{M}\boldsymbol{l} = 0, \tag{3.4d}$$

$$\boldsymbol{r}'\cdot\boldsymbol{d}_3 - 1 = 0, \tag{3.4e}$$

$$\boldsymbol{r}'\wedge\boldsymbol{d}_3 + \Upsilon\boldsymbol{d}_3 = \mathbf{0}, \tag{3.4f}$$

$$\boldsymbol{n}^L\cdot\boldsymbol{d}_3 = 0, \tag{3.4g}$$

where

$$\boldsymbol{n} := -\boldsymbol{n}^L\wedge(\mathtt{R}\boldsymbol{d}_3^{\dagger}) + n_{||}\mathtt{R}\boldsymbol{d}_3^{\dagger}, \tag{3.5a}$$

$$\boldsymbol{m} := \boldsymbol{m}_{\perp} + \boldsymbol{m}_{||}, \tag{3.5b}$$

$$\boldsymbol{m}_{\perp} := \beta\mathtt{R}(\boldsymbol{d}_3^{\dagger}\wedge\mathtt{R}^{\top}\mathtt{R}'\boldsymbol{d}_3^{\dagger}), \tag{3.5c}$$

$$\text{and } \boldsymbol{m}_{||} := \gamma\boldsymbol{d}_2^{\dagger}\cdot(\mathtt{R}^{\top}\mathtt{R}'\boldsymbol{d}_1^{\dagger})\mathtt{R}\boldsymbol{d}_3^{\dagger}. \tag{3.5d}$$

In above Eqs. (3.1), (3.4), $\nabla$ denotes the differential operator (gradient) with respect to $\boldsymbol{x}\in\Omega$ and $()'$ denotes the ordinary derivative of $()$ with respect to the arclength parameter of $\Gamma$, i.e. $s\in\mathcal{I}$. In Eq. (3.4), the derivatives of the unit normal, $\boldsymbol{l}$ and the unit tangent $\boldsymbol{t}$ of a circle of diameter $\mathfrak{D}$, with respect to the arclength parameter satisfy the equations:

$$\boldsymbol{t}' = -\frac{2}{\mathfrak{D}}\boldsymbol{l}, \quad \boldsymbol{l}' = \frac{2}{\mathfrak{D}}\boldsymbol{t}. \tag{3.6}$$

Also, in above Eq. (3.3), $\mathbf{I}$ is identity tensor on two dimensions (i.e., on span$(\boldsymbol{e}_1, \boldsymbol{e}_2)$) and in (3.4e)-(3.4g), by (1.14), $\boldsymbol{d}_3 = \mathtt{R}\boldsymbol{d}_3^{\dagger}$. From the perspective of generalized non-linear elasticity $\widetilde{\mathbf{N}} : \widetilde{\Omega} \to \mathrm{Sym}$ (3.3a) and $\widetilde{\mathbf{M}} : \widetilde{\Omega} \to \mathrm{Sym}$ (3.3b) represent the referential (Piola-Kirchhoff) stress tensor field and the referential couple stress tensor field of the plate, respectively. See Fig. 3 for the mechanical free body diagram of an element in the interior of plate (without the decoration $\tilde{}$). In the physical structure of the mechanical system, $\boldsymbol{m}$ is the moment at a point with $\boldsymbol{m}_\perp$ and $\boldsymbol{m}_{||}$ representing the bending moment and the twisting moment, respectively, at the material point $\boldsymbol{x}$ of the rod while $\boldsymbol{n}$ (see (3.5a) below) is the force at that point. See Fig. 4 for the mechanical free body diagram of an element at the boundary of plate (without the decoration $\tilde{}$). Based on the statement immediately following (2.6), note that $\boldsymbol{n}$ has the units of $[N]$, $\boldsymbol{m}$ has the units of $[N][m]$, whereas $\widetilde{\mathbf{N}}$, $\mathbf{N}$ have units $[N]/[m]$, but $\widetilde{\mathbf{M}}$, $\mathbf{M}$ have the units $[N]$.

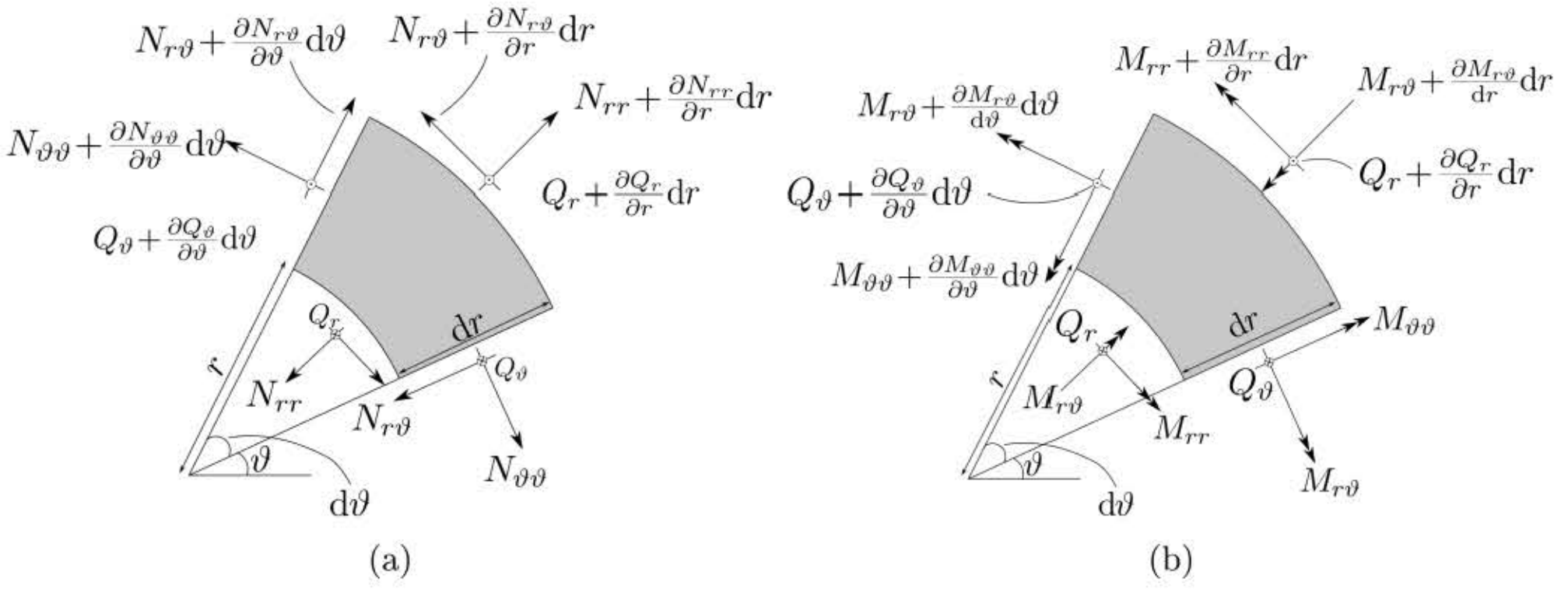

Figure 3: (a) Force based and (b) moment based in-plane and out-of-plane stress components for interior (domain) element. $\odot, \otimes$ indicate out-of-plane $+, -$ components (along $\boldsymbol{e}_3$ axis).

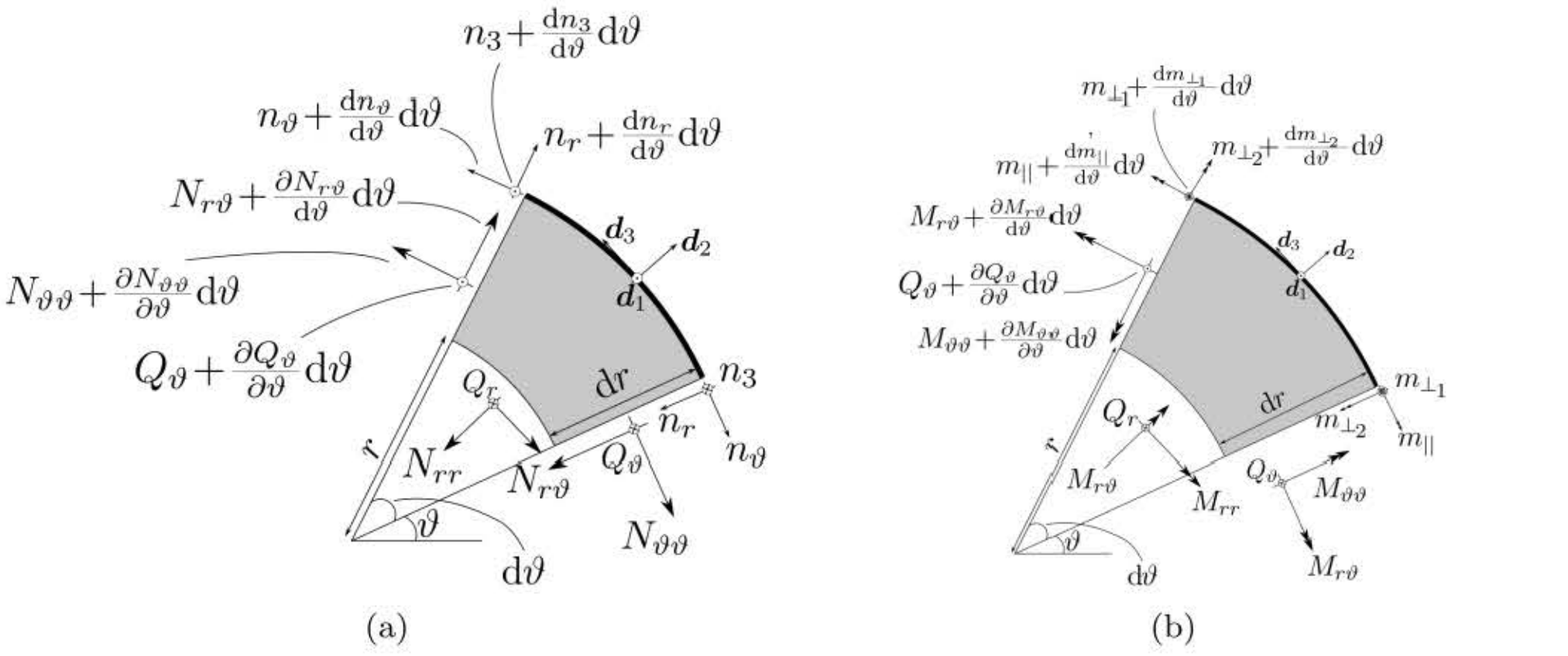

Figure 4: (a) Force based and (b) moment based in-plane and out-of-plane stress components for edge (boundary) element. $\odot, \otimes$ indicate out-of-plane $+, -$ components (along $\boldsymbol{e}_3$ axis).

**Remark 3.** *There is a clear distinction in the definition of* $\mathbf{M}$ *in* $(3.2)_2$ *with respect to that of* $\mathbf{N}$ *in* $(3.2)_1$*. This is justified later in the article in connection with the desirable form of the bifurcation problem formulation.*

**Remark 4.** *Fig. 3a–Fig. 4b are the schematic free body diagrams of infinitesimal element of the plate. In these figures,* $\widetilde{\mathbf{N}}$ *is written as* $\mathbf{N}$ *and* $\widetilde{\mathbf{M}}$ *as* $\mathbf{M}$*. The figures are partially* *based on the discussion that appears in Mittelstedt (2023). In terms of the components, using polar coordinates, the tensors* $\mathbf{N}$ *and* $\mathbf{M}$ *can be expressed as*

$$\mathbf{N} = N_{rr}\boldsymbol{e}_r \otimes \boldsymbol{e}_r + N_{\theta\theta}\boldsymbol{e}_\theta \otimes \boldsymbol{e}_\theta + N_{r\theta}(\boldsymbol{e}_r \otimes \boldsymbol{e}_\theta + \boldsymbol{e}_\theta \otimes \boldsymbol{e}_r), \tag{3.7}$$

$$\mathbf{M} = M_{rr}\boldsymbol{e}_r \otimes \boldsymbol{e}_r + M_{\theta\theta}\boldsymbol{e}_\theta \otimes \boldsymbol{e}_\theta + M_{r\theta}(\boldsymbol{e}_r \otimes \boldsymbol{e}_\theta + \boldsymbol{e}_\theta \otimes \boldsymbol{e}_r), \tag{3.8}$$

*where* $\boldsymbol{e}_r, \boldsymbol{e}_\theta$ *are defined in* (1.10).

*Fig. 3a and Fig. 3b are corresponding to interior element, and Fig. 4a and Fig. 4b are corresponding to an element adjacent to the edge. Fig. 3a shows the in-plane stress* $N_{ij}, i, j \in \{r, \theta\}$ *and out of plane shear stress* $Q_i, i \in \{r, \theta\}$ *in polar components. Fig. 3b shows the couple stress* $M_{ij}, i, j \in \{r, \theta\}$ *along with the out of plane shear* $Q_i$*. The term* $Q_i$ *can be eliminated by first performing moment balance along* $\boldsymbol{e}_\theta$ *and* $\boldsymbol{e}_r$ *directions to obtain* $Q_i$ *in terms of derivatives of the couple stress components,* $M_{ij}$*, and then substituting them in force balance equation along* $e_3$ *direction to obtain the equilibrium equations. Due to this reason,* $Q_i$ *doesn't appear in the final form of the out of plane equilibrium equation,* (3.1b)*. The in-plane force equilibrium equations along* $\boldsymbol{e}_r$ *and* $\boldsymbol{e}_\theta$ *are completely in terms of* $N_{ij}$ *and its derivative as obtained in* (3.1a)*. The moment balance in* $\boldsymbol{e}_3$ *direction gives the symmetry condition of in-plane stress tensor* $\mathbf{N}$*. Similarly, the domain equations for the the element closer to edge, as shown in Fig. 4a and Fig. 4b, are found to be the same as that of interior element. In addition, we get additional equations on the boundary of the plate. Moment balance of the edge rod element gives the vector valued equation,* (3.4a)*. In plane force balance and out of plane force balance of the rod, results in the equations* (3.4b) *and* (3.4c)*, respectively. The moment balance in* $\boldsymbol{e}_\theta$ *direction of the plate element attached to the rod, gives the equation* (3.4d) *as the plate has a hinge-like connection to the rod. The equations* (3.4e)-(3.4g) *are the inextensibility and unshearability constraints on the rod.*

Taking the dot product of (3.4f) with $\boldsymbol{d}_3$ gives $\Upsilon = 0$ (recall $(2.3)_2$, (2.6)). Hence, (3.4a) is simplified to

$$\boldsymbol{m}' + (\boldsymbol{n}^L \wedge \boldsymbol{d}_3) \wedge \boldsymbol{r}' = \boldsymbol{0}, \tag{3.9}$$

and (3.4f) simplifies to (1.19). Using the inextensibility constraint (3.4e), the unshearability constraint (1.19), and the constraint (3.4g), it is found that

$$\begin{aligned}(\boldsymbol{n}^L \wedge \boldsymbol{d}_3) \wedge \boldsymbol{r}' &= -\,\boldsymbol{n} \wedge \boldsymbol{r}', \\ n_{||} &= \boldsymbol{n} \cdot \boldsymbol{d}_3,\end{aligned} \tag{3.10}$$

where $\boldsymbol{n}$ is defined in (3.5a). Taking the cross product of (3.5a) with $\boldsymbol{d}_3$ and using (3.4g) gives $\boldsymbol{n}^L = \boldsymbol{n} \wedge \boldsymbol{d}_3$, so that (3.4g) becomes redundant. Instead of a vector field $\boldsymbol{n}^L$ and a scalar field $n_{||}$, the equations can be simply expressed in terms of a single vector field $\boldsymbol{n}$ (3.5a). Thus, the constraint equations (3.4e) and (3.4f) can be replaced by

$$\boldsymbol{r}' - \boldsymbol{d}_3 = 0. \tag{3.11}$$

As $\mathtt{R}(s) \in \mathrm{SO}(3)$ at each $s \in \mathcal{I}$, the arclength parameter of $\mathcal{I}$, we express it in terms of its Gibbs vector so that

$$\mathtt{R}(s) := \mathrm{Rot}(\boldsymbol{g}(s)), \tag{3.12}$$

at each $s \in \mathcal{I}$, where

$$\mathrm{Rot}(\boldsymbol{g}) := \frac{1}{1 + \boldsymbol{g} \cdot \boldsymbol{g}}((1 - \boldsymbol{g} \cdot \boldsymbol{g})\mathtt{I} + 2\boldsymbol{g} \otimes \boldsymbol{g} + 2\mathrm{skw}(\boldsymbol{g})), \tag{3.13}$$

and the notation skw($\mathbf{a}$) denotes the skew tensor corresponding to a given vector $\mathbf{a} \in \mathbb{R}^3$. Above description (3.12) suffices as we are investigating locally, i.e. in the neigbourhood of a configuration. We utilized the property of the map Rot (3.13) that it is locally invertible, i.e. invertible in such neigbourhood.

## 4. Bifurcation from homogeneously deformed plate: consequences of necessary condition

It is easy to verify that the system of partial differential equations (3.1) with boundary condition (3.4), along with subsidiary conditions (3.2), (3.3a), (3.3b), and (3.5c), (3.5d), has the following solution for all $\lambda \in \mathbb{R}^+$, referred as the base (principal or trivial) solution:

$$\boldsymbol{v}(\boldsymbol{x}) = \mathbf{0},\ w(\boldsymbol{x}) = 0,\ \boldsymbol{g}(s) = \mathbf{0},\ \boldsymbol{n}(s) = 6\mathfrak{D}(1+\nu)(1-\lambda)(-1)\boldsymbol{t}(s), \tag{4.1}$$

where $\boldsymbol{x} \in \Omega$ and $s \in \mathcal{I}$.

There is an explicit presence of the bifurcation parameter $\lambda$ in the description of this base solution branch (4.1). Recall the statement following (1.22). In fact, the formulation presented so far can be also written with explicit presence of many different structure parameters assumed in the model so that, in principle, there are several other potential choices of parameters which may lead to notable changes in the behaviour of the base solution branch. However, as mentioned just after (1.22), from physical perspective, $\lambda$ plays a key role, so we regard $\lambda$ as the bifurcation parameter.

In order to carry out the local bifurcation analysis of the above mentioned base solution branch, the formulation, presented so far, is conveniently re-wrtitten in an operator form. This requires an introduction of some more fields and is described below.

### *4.1. Operator form of equations of equilibrium*

Let $(r, \theta)$ represent the polar co-ordinates of a point $\boldsymbol{x} \in \Omega$. The polar basis vectors $\boldsymbol{e}_r, \boldsymbol{e}_\theta$ are defined in (1.10). The stress tensor $\mathbf{N}$ and moment tensor $\mathbf{M}$, in view of (3.2), are written in the the polar co-ordinates in the same way as in (3.7), (3.8), respectively. Consider the definition

$$\boldsymbol{f} := \frac{1}{\lambda}\mathbf{N}\nabla w. \tag{4.2}$$

The collection of all relevant fields, according to their definitions provided earlier in the mechanics formulation §1.1, §1.2, §1.3, §3, and, in particular, (3.2), is represented by

$$\begin{aligned}\boldsymbol{U} :=& \Big(\boldsymbol{v}, w, \mathbf{N}, \mathbf{M}, \boldsymbol{f}, \boldsymbol{g}, \boldsymbol{n}\Big)^{\top} \\ =& \Big(\boldsymbol{v}, w, \big[N_{rr}, N_{\theta\theta}, N_{r\theta}\big], \big[M_{rr}, M_{\theta\theta}, M_{r\theta}\big], \boldsymbol{f}, \boldsymbol{g}, \boldsymbol{n}\Big)^{\top} \in \mathcal{U},\end{aligned} \tag{4.3}$$

where

$$\mathcal{U} := (\mathcal{C}^1)^2 \times \mathcal{C}^2 \times \mathcal{C}^1 \times \mathcal{C}^1 \times \mathcal{C}^1 \times \mathcal{C}^2 \times \mathcal{C}^2 \times \mathcal{C}^2 \times (\mathcal{C}^1)^2 \times (\mathcal{P}^2)^3 \times (\mathcal{P}^1)^3. \tag{4.4}$$

Note that $\Gamma = \partial\Omega$ (1.25), which along with (3.3a) and (3.3b) leads to the equations of equilibrium within the plate and the boundary conditions as stated in the Appendix B. In above, $\mathcal{P}^k$ is the set of $k$ times continuously differentiable periodic functions defined on $\Gamma$ and $\mathcal{C}^k$ is the set of $k$ times continuously differentiable functions defined on the closure of $\Omega$. The component-wise distribution of units in $\boldsymbol{U}$ is $\Big([m], [m], [N]/[m], [N], [N]/[m], [m]^0, [N]\Big)^{\top}$.

The base solution (4.1) can be expressed (in terms of the newly introduced notation $\boldsymbol{U}$ in (4.3)) as $\boldsymbol{U}_0$ where

$$\boldsymbol{U}_0 := 12\mathfrak{C}(1+\nu)(1-\lambda) \times \Big(\mathbf{0}, 0, \big[1,1,0\big], \big[0,0,0\big], \mathbf{0}, \mathbf{0}, \frac{1}{2}\mathfrak{D}(-1)\boldsymbol{t}\Big)^{\top}, \tag{4.5}$$

where $\mathfrak{C}$ is defined in (1.8). In order to study the perturbations emanating off the base solution, let

$$\boldsymbol{V} := \boldsymbol{U} - \boldsymbol{U}_0, \tag{4.6}$$

$$\text{with } \boldsymbol{V} := \Big(\mathring{\boldsymbol{v}}, \mathring{z}, \big[\mathring{N}_{rr}, \mathring{N}_{\theta\theta}, \mathring{N}_{r\theta}\big], \big[\mathring{M}_{rr}, \mathring{M}_{\theta\theta}, \mathring{M}_{r\theta}\big], \mathring{\boldsymbol{f}}, \mathring{\boldsymbol{g}}, \mathring{\boldsymbol{n}}\Big)^{\top}, \tag{4.7}$$

so that the base solution is indeed the zero (trivial) solution relative to the space of perturbations. In this context, the corresponding tensor perturbations $\mathring{\mathbf{N}}, \mathring{\mathbf{M}}$ are defined in the same way as in (3.7), (3.8), respectively. The equations from Appendix B can be re-written and are provided in Appendix C.

**Remark 5.** *A few comments regarding Appendix C are useful in order to proceed further with a clear separation of equations of equilibrium into those which capture the important role, such as non-linear terms, vs those which can be included in a suitable function space. Note that the collection of equations in (C.2) are related to a linear operator while (C.3) is related to a non-linear operator. We identify (C.2) as the linear boundary equations and (C.3) as the non-linear boundary equations. In connection with Remark 3, it is evident that the definition of* $\mathbf{M}$ *in (3.2) permits (C.2) to be independent of the bifurcation parameter* $\lambda$.

For convenience of notation, the decoration $\mathring{}$ is dropped in the context of working further with the equations listed in Appendix C.

In order to proceed further with slightly abstract formulation, it is useful to recall (4.4). Let

$$\begin{aligned}\mathcal{D} =& \{\boldsymbol{V} \in \mathcal{U} : \text{(C.2) is satisfied}\},\\ \mathcal{C} =& (\mathcal{C}^0)^2 \times \mathcal{C}^0 \times \mathcal{C}^0 \times \mathcal{C}^0 \times \mathcal{C}^0 \times \mathcal{C}^0 \times \mathcal{C}^0 \times \mathcal{C}^0 \times (\mathcal{C}^1)^2 \times (\mathcal{P}^0)^3 \times (\mathcal{P}^0)^3.\end{aligned} \tag{4.8}$$

By virtue of the observation in previous Remark 5 regarding linearity of (C.2) and independence from the bifurcation parameter $\lambda$, we emphasize that $\mathcal{D}$ is a linear space.

Let

$$\mathfrak{F} : \mathcal{D} \times \mathbb{R}^+ \to \mathcal{C} \tag{4.9}$$

be defined such that

$$\mathfrak{F}(\boldsymbol{V}, \lambda) = \left(\begin{array}{c} \nabla \cdot \mathbf{N} \\ \nabla \cdot (\boldsymbol{f} + \nabla \cdot \mathbf{M}) \\ 6\mathfrak{C}\left((1-\nu)\big(\nabla \boldsymbol{v} + \nabla \boldsymbol{v}^{\top} + \frac{1}{\lambda}\nabla z \otimes \nabla z\big) + \nu\big(2\nabla \cdot \boldsymbol{v} + \frac{1}{\lambda}\nabla z \cdot \nabla z\big)\mathbf{I}\right) - \mathbf{N} \\ -\mathfrak{C}\frac{h^2}{\lambda^2}((1-\nu)\nabla^2 z + \nu \Delta z \mathbf{I}) - \mathbf{M} \\ \boldsymbol{f} - \frac{12}{\lambda}\mathfrak{C}(1+\nu)(1-\lambda)\nabla z - \frac{1}{\lambda}\mathbf{N}\nabla z \\ \boldsymbol{m}' - (\boldsymbol{n} + 6\mathfrak{C}\,\mathfrak{D}(1+\nu)(\lambda-1)\boldsymbol{t}) \wedge (\boldsymbol{t} + \boldsymbol{v}' + z'\boldsymbol{e}_3) \\ \boldsymbol{t} + \boldsymbol{v}' + z'\boldsymbol{e}_3 - \boldsymbol{d}_3 \end{array}\right), \tag{4.10}$$

where the definitions of $\boldsymbol{m}, \boldsymbol{m}_{\perp}$ and $\boldsymbol{m}_{||}$ are provided in (3.5b), (3.5c) and (3.5d), respectively, after substitution of (3.12).

The components of $\mathfrak{F}(\boldsymbol{V}, \lambda)$ (4.10) are given by the left hand side of (C.1) and (C.3).

As mentioned before, for convenience, for example, we also identify $\boldsymbol{v}(\boldsymbol{x}(s))$ as $\underline{\boldsymbol{v}}(s)$ and simply write it as $\boldsymbol{v}(s)$, and in a similar way several other fields defined on the plate but restricted to the boundary, i.e the rod; In other words, $\boldsymbol{v}(s) \equiv \underline{\boldsymbol{v}}(s) = \boldsymbol{v}(\boldsymbol{x}(s))$, and $z(s) \equiv \underline{z}(s) = z(\boldsymbol{x}(s))$, for all $s \in \mathcal{I}$.

The nonlinear problem of finding the equilibrium configuration of the rod-plate mechanical system is represented by

$$\mathfrak{F}(\boldsymbol{V}, \lambda) = \mathbf{0} \text{ where } \boldsymbol{V} \in \mathcal{D},\ \lambda \in \mathbb{R}^+. \tag{4.11}$$

The base solution (4.1), that is (4.5), owing to the change of variables (4.6), corresponds to the trivial solution of (4.11) (now the base solution of (4.11)) is given by

$$(\boldsymbol{V}, \lambda) = (\mathbf{0}, \lambda) \text{ for all } \lambda \in \mathbb{R}^+. \tag{4.12}$$

Before continuing the analysis further, it is useful to define certain length scales, dimensionless structure parameters, and dimensionless fields.

*4.2. Dimensionless formulation*

We can use many choices to define the length scale $L$, for example, $L := \frac{1}{2}\mathfrak{D}, L := (\beta/\mathfrak{C})^{1/3}, \ldots$ . For the second choice, in the subsequent sections, we need to assume that the rod has non-zero bending stiffness. The dimensionless co-ordinates and arclengths are scaled by the length scale $L$ and are given by

$$\boldsymbol{x}^* := \frac{\boldsymbol{x}}{L}, \quad s^* := \frac{s}{L}. \tag{4.13a}$$

Additionally, we define the following dimensionless structure parameters:

$$h^* := \frac{h}{L}, \quad \mathfrak{D}^* := \frac{\mathfrak{D}}{L}, \quad \alpha := \frac{2L}{\mathfrak{D}}, \quad \beta^* := \frac{\beta}{\mathfrak{C}L^3}, \quad \gamma^* := \frac{\gamma}{\mathfrak{C}L^3}. \tag{4.13b}$$

We define the following dimensionless fields and variables, in the context of the original symbols utilized in the definition (4.7):

$$\mathring{\boldsymbol{v}}^* := \frac{\mathring{\boldsymbol{v}}}{L}, \quad \mathring{z}^* := \frac{\mathring{z}}{L}, \quad \mathring{\mathbf{N}}^* := \frac{\mathring{\mathbf{N}}}{\mathfrak{C}}, \quad \mathring{\mathbf{M}}^* := \frac{\mathring{\mathbf{M}}}{\mathfrak{C}L}, \tag{4.13c}$$

$$\mathring{\boldsymbol{f}}^* := \frac{\mathring{\boldsymbol{f}}}{\mathfrak{C}}, \quad \mathring{\boldsymbol{g}}^* := \mathring{\boldsymbol{g}}, \quad \mathring{\boldsymbol{n}}^* := \frac{\mathring{\boldsymbol{n}}}{\mathfrak{C}L}. \tag{4.13d}$$

That the above set of definitions is indeed a recipee for dimensionless formulation is not surprising in view of the statement immediately following (4.4) and noting that the unit of $L$ is $[m]$ and the units of $\mathfrak{C}$ in (1.8) are $[N]/[m]$.

Above dimensionless quantities are substituted in (C.1), (C.2) and (C.3) and the equation-wise common factors are pulled out to obtain dimensionless form $\mathfrak{F}^*$ of $\mathfrak{F}$ (4.10).

Naturally, the choice of $L = \frac{1}{2}\mathfrak{D}$ is equivalent to substituting $\mathfrak{D}^* = 2$ that is the rod is a unit circle in its stress-free configuration in the dimensionless formulation, whereas the choice of $L = (\beta/\mathfrak{C})^{1/3}$ is equivalent to substituting the rod bending modulus $\beta^* = 1$. In this article, we choose the former so that the scaled rod center-line in the rod's stress-free configuration is the unit circle $\mathbb{T}$ in the plane $\mathbb{R}^2$. However, for clarity and to keep connection with the physical problem, we do not remove $\mathfrak{D}^*$ from the ensuing mathematical expressions and relevant equations.

The decoration $*$ is dropped in the following and the dimensionless quantities are denoted without any decorations. Recall that the decoration $^\circ$ is also dropped for convenience, as already mentioned.

*4.3. Linearization of equilibrium equations relative to base solution*

In this section, the linearization of (4.11) about $\boldsymbol{V}=0$ along

$$\boldsymbol{X}=\Big(\boldsymbol{v}, z, \mathbf{N}, \mathbf{M}, \boldsymbol{f}, \boldsymbol{s}, \boldsymbol{n}\Big)^{\top} \in \mathcal{D} \tag{4.14}$$

is considered, i.e. $\boldsymbol{V}=\mathbf{0}+\varepsilon\boldsymbol{X}+o(\varepsilon)$ as $\varepsilon\to 0$.

At this point it is useful to recall the definitions (4.6), (4.7) as well as the comment regarding the notation in the sentences following these equations. The linearization of the rotation tensor field using (3.12) is given by

$$\mathtt{R}=\mathtt{I}+2\varepsilon\ \mathrm{skw}(\boldsymbol{s})+o(\varepsilon),\ \text{ as } \varepsilon\to 0. \tag{4.15}$$

In the contex of (A.5), recall that the unit normal and unit tangent vectors at any $\boldsymbol{x}\in\Gamma$ are given by $\boldsymbol{l}(\boldsymbol{x})=\boldsymbol{x}/\sqrt{\boldsymbol{x}\cdot\boldsymbol{x}}, \boldsymbol{t}(\boldsymbol{x})=\boldsymbol{e}_3\wedge\boldsymbol{l}(\boldsymbol{x})$. Using (3.5c), (3.5d) and (4.15), the linearization of moments in the rod is given by

$$\boldsymbol{m}_{\perp}=2\varepsilon\ \beta(\mathtt{I}-\boldsymbol{t}\otimes\boldsymbol{t})\boldsymbol{s}'+o(\varepsilon),\quad \boldsymbol{m}_{||}=2\varepsilon\ \gamma(\boldsymbol{t}\otimes\boldsymbol{t})\boldsymbol{s}'+o(\varepsilon),\ \text{ as } \varepsilon\to 0. \tag{4.16}$$

Differentiating (1.26) with respect to arclength parameter $s$, using the fact that $\widetilde{\boldsymbol{r}}(\widetilde{s})=\boldsymbol{r}(s)$ and substituting the linearization of the fields gives

$$\boldsymbol{r}'(s)=\boldsymbol{t}(s)+\varepsilon\ \frac{\mathrm{d}}{\mathrm{d}s}\boldsymbol{v}(\boldsymbol{x}(s))+\varepsilon\ \boldsymbol{e}_3\frac{\mathrm{d}}{\mathrm{d}s}z(\boldsymbol{x}(s))+o(\varepsilon),\ \text{ as } \varepsilon\to 0, \tag{4.17}$$

for all $s\in\mathcal{I}$. Recall that $()'$ denotes the ordinary derivative of $()$ with respect to the arclength parameter and we suppress the presence of the arclength parameter.

Using (4.16) and (4.17) to linearize $(\mathrm{C.3})_1$, i.e. $\boldsymbol{m}'-(\varepsilon\ \boldsymbol{n}+\frac{12}{\alpha}(1+\nu)(\lambda-1)\boldsymbol{t})\wedge\boldsymbol{r}'=\mathbf{0}$, gives

$$\begin{aligned}&(2\beta\mathtt{I}+2(\gamma-\beta)\boldsymbol{t}\otimes\boldsymbol{t})\boldsymbol{s}''-4\frac{(\gamma-1)}{\mathfrak{D}}(\boldsymbol{t}\otimes\boldsymbol{l}+\boldsymbol{l}\otimes\boldsymbol{t})\boldsymbol{s}'+\\&+\boldsymbol{t}\wedge(\boldsymbol{n}-\frac{12}{\alpha}(1+\nu)(\lambda-1)(\boldsymbol{v}'+z'\boldsymbol{e}_3))=o(1),\ \text{ as } \varepsilon\to 0.\end{aligned} \tag{4.18}$$

Using (1.14), as a result of (4.15), the linearization of $\boldsymbol{d}_3$ is given by

$$\boldsymbol{d}_3=\boldsymbol{t}+2\varepsilon\ \boldsymbol{s}\wedge\boldsymbol{t}+o(\varepsilon),\ \text{ as } \varepsilon\to 0. \tag{4.19}$$

Substituting (4.19) and (4.17) in $(\mathrm{C.3})_2$, i.e. in $\boldsymbol{r}'-\boldsymbol{d}_3=\mathbf{0}$, we get

$$\boldsymbol{v}'+z'\boldsymbol{e}_3-2\boldsymbol{s}\wedge\boldsymbol{t}=o(1),\ \text{ as } \varepsilon\to 0. \tag{4.20}$$

The equations (C.2) are satisfied by definition (4.8). The linearization of (C.1) leads to dropping of second order terms from $(\mathrm{C.1})_3$, $(\mathrm{C.1})_4$, $(\mathrm{C.1})_5$ and $(\mathrm{C.1})_8$.

The linearization of the operator $\mathfrak{F}$ at $\boldsymbol{V}=\mathbf{0}$ is denoted by $\mathfrak{L}(\lambda)$.

With $\boldsymbol{V}=\mathbf{0}+\varepsilon\boldsymbol{X}+o(\varepsilon)$ as $\varepsilon\to 0$ and $\boldsymbol{X}$ as in (4.14), $\mathfrak{L}(\lambda)$ is, essentially, the Fréchét derivative $\mathrm{D}_{\boldsymbol{V}}\mathfrak{F}(\boldsymbol{V},\lambda)|_{\boldsymbol{V}=\mathbf{0}}$, relative to its first argument $\boldsymbol{V}$ (Kielhöfer, 2006).

The expression for $\mathfrak{L}(\lambda)$ is found to be

$$\mathfrak{L}(\lambda)[\boldsymbol{X}]=\begin{pmatrix}\nabla\cdot\mathbf{N}\\ \nabla\cdot(\boldsymbol{f}+\nabla\cdot\mathbf{M})\\ 6\left((1-\nu)\left(\nabla\boldsymbol{v}+\nabla\boldsymbol{v}^{\top}\right)+2\nu\nabla\cdot\boldsymbol{v}\mathbf{I}\right)-\mathbf{N}\\ -\frac{h^2}{\lambda^2}((1-\nu)\nabla^2 z+\nu\Delta z\mathbf{I})-\mathbf{M}\\ \boldsymbol{f}-\frac{12}{\lambda}(1+\nu)(1-\lambda)\nabla z\\ (2\beta\mathtt{I}+2(\gamma-\beta)\boldsymbol{t}\otimes\boldsymbol{t})[\boldsymbol{s}'']-2\alpha(\gamma-1)(\boldsymbol{t}\otimes\boldsymbol{l}+\boldsymbol{l}\otimes\boldsymbol{t})[\boldsymbol{s}']+\dots\\ \dots+\boldsymbol{t}\wedge\left(\boldsymbol{n}-\frac{12}{\alpha}(1+\nu)(\lambda-1)(\boldsymbol{v}'+z'\boldsymbol{e}_3)\right)\\ \boldsymbol{v}'+z'\boldsymbol{e}_3-2\boldsymbol{s}\wedge\boldsymbol{t}\end{pmatrix}, \tag{4.21}$$

for all $\boldsymbol{X} \in \mathcal{D}$, with component variables defined by (4.14).

A necessary condition for a bifurcation point in the problem (4.11), along the trivial solution branch, is that the linear operator $\mathfrak{L}(\lambda)$ is singular (Kielhöfer, 2006). In other words, in such a case that particular value of $\lambda$ is a critical value which is a candidate point for a bifurcation relative the trivial solution branch. In the following section, the critical values of $\lambda \in \mathbb{R}^+$ and the null space $\{\boldsymbol{X}_\lambda\} \subset \mathcal{D}$ of the linear operator $\mathfrak{L}(\lambda)$ are determined, i.e, $(\lambda, \boldsymbol{X}_\lambda)$ such that

$$\mathfrak{L}(\lambda)[\boldsymbol{X}_\lambda] = \mathbf{0}, \text{where } \boldsymbol{X}_\lambda \neq \mathbf{0}. \tag{4.22}$$

*4.4. Formulation using polar co-ordinates*

The surface $\Omega$ is parametrized using polar co-ordinates $(r, \theta) \in [0, \mathfrak{D}/2) \times [0, 2\pi) = \mathcal{S}$. Under this parametrization, any point $\boldsymbol{x} \in \Omega$ is given by

$$\boldsymbol{x} = r\cos\theta \boldsymbol{e}_1 + r\sin\theta \boldsymbol{e}_2. \tag{4.23}$$

Similarly, the boundary $\partial\Omega$ is parametrized using $\theta \in [0, 2\pi) = \mathcal{L}$. The arclength parameter of the rod, $s$, in terms of $\theta$ is

$$s = \frac{\theta}{\alpha}. \tag{4.24}$$

Using (4.23) and (4.24), all the fields on $\Omega$ and $\partial\Omega$ are mapped on $\mathcal{S}$ and $\mathcal{L}$ respectively.

The vector valued terms of $\boldsymbol{X}_\lambda \equiv \boldsymbol{X}(r, \theta)$, namely $\boldsymbol{v}$, $\boldsymbol{f}$, $\boldsymbol{s}$ and $\boldsymbol{n}$ are expressed in polar co-ordinates as follows:

$$\begin{aligned} \boldsymbol{v} = v_r\boldsymbol{e}_r + v_\theta\boldsymbol{e}_\theta, \quad \boldsymbol{f} = f_r\boldsymbol{e}_r + f_\theta\boldsymbol{e}_\theta, \quad \boldsymbol{s} &= s_r\boldsymbol{e}_r + s_\theta\boldsymbol{e}_\theta + s_3\boldsymbol{e}_3, \\ \boldsymbol{n} &= n_r\boldsymbol{e}_r + n_\theta\boldsymbol{e}_\theta + n_3\boldsymbol{e}_3. \end{aligned} \tag{4.25}$$

It noted that $\boldsymbol{e}_r = \boldsymbol{l}$ and $\boldsymbol{e}_\theta = \boldsymbol{t}$.

It is noted that the linearized problems, both the domain equations and boundary conditions, involving the in-plane displacement $(v_r, v_\theta)$ and the out-of-plane displacement $z$ are decoupled. The details of the formulation using polar co-ordinates are available in Appendix D. Hence, in the following, the corresponding in-plane solution (null-space of first kind) and the out-of-plane solution (null-space of second kind) are stated independent of each other.

*4.5. In-plane perturbed solution relative to base solution*

As the intermediate configuration (the configuration of trivial branch) of rod-plate systrem is a circular disc, the displacement fields are periodic with respect to variable $\theta$. The planar displacement components are written as Fourier series in $\theta$. The following form of solution for $v_r$ and $v_\theta$ holds:

$$v_r(r, \theta) = \sum_{k=0}^{\infty} a_k(r)\cos k\theta + \sum_{k=1}^{\infty} b_k(r)\sin k\theta, \tag{4.26a}$$

$$v_\theta(r, \theta) = \sum_{k=0}^{\infty} c_k(r)\cos k\theta + \sum_{k=1}^{\infty} d_k(r)\sin k\theta. \tag{4.26b}$$

The expressions (4.26) are substituted into the system of partial differential equations (D.1) and the coefficients of sine and cosine terms are collected to form a system of ordinary differential equations of $r$. The system of ordinary differential equations is solved to obtain functions $a_k$, $b_k$, $c_k$ and $d_k$. As the displacements should be finite at $r=0$, the following general solution of (D.1) is obtained:

$$\begin{aligned} v_r(r,\theta) = \quad & A_0 r + \sum_{k=1}^{\infty}\left(-A_k r^{k-1} - C_k \tfrac{k-2+\nu(k+2)}{4+k(1+\nu)} r^{k+1}\right)\cos k\theta + \\ & + \sum_{k=1}^{\infty}\left(B_k r^{k-1} + D_k \tfrac{k-2+\nu(k+2)}{4+k(1+\nu)} r^{k+1}\right)\sin k\theta, \qquad (4.27a) \\ v_\theta(r,\theta) = \quad & B_0 r + \sum_{k=1}^{\infty}\left(B_k r^{k-1} + D_k r^{k+1}\right)\cos k\theta + \\ & + \sum_{k=1}^{\infty}\left(A_k r^{k-1} + C_k r^{k+1}\right)\sin k\theta. \qquad (4.27b) \end{aligned}$$

The critical values of $\lambda$ are obtained by substituting (4.27) in boundary conditions (D.9) and (D.4e) and solving for non-trivial set of coefficient $A_k$, $B_k$, $C_k$ and $D_k$.

As the problem is linear and the general solutions for each value $k$ are linearly independent of each other, we insist that the boundary conditions are satisfied for each value of $k$ separately. The critical value corresponding to each value of $k$ is denoted by $\lambda_k$.

It is found that (D.9) and (D.4e) are satisfied for all values of $\lambda$ for $k=0$ and 1. These solutions correspond to rigid rotation about $\boldsymbol{e}_3$ axis and rigid planar translation, respectively. The corresponding coefficients, $A_0 = 0$, $C_1 = D_1 = 0$, and, $B_0$, $A_1$ and $B_1$ are arbitrary. For each $k \geq 2$, the critical values of lambda are found to be,

$$\lambda_k = \frac{\begin{array}{c} 48 + k(24-24\nu) + \beta\alpha^3 k^4(3-\nu) \\ +k^2(-36 - 3\beta\alpha^3 - 24\nu + \beta\alpha^3\nu + 12\nu^2) \end{array}}{12(\nu^2-2\nu-3)(k^2-1)}. \qquad (4.28)$$

Corresponding to each $k \geq 2$, the relation between the coefficients $A_k$, $B_k$, $C_k$ and $D_k$ is

$$A_k : C_k = B_k : D_k = -\frac{(k+1)(-2\nu+\nu k+k+2)}{\alpha^2(k-1)(\nu k+k+4)} : 1. \qquad (4.29)$$

**Remark 6.** *The expression* (4.28) *is valid only for* $\lambda_k > 0$. *Note that, as* $k \to \infty$, $\lambda_k \to -\infty$. *As a consequence, the range of k is finite, i.e., there are only a finite number of critical point* $\lambda_k$, *where* $k = 2, \ldots N_c$. $N_c$ *depends on* $\nu$ *and* $\alpha$, *and is finite because the rod is under compression in the presence of a bounded contact force (with an upper limit) provided by the plate which prohibits highly oscillatory instability modes (the plate model is pathological in the extreme case of deformation when the circular plate is shrunk to a point, as this process can be carried out with finite radial traction). Also when the plate is very soft, the set of such* $\lambda_k$ *can be empty.*

From above calculations, we find that $\mathfrak{L}(\lambda)$ has two dimensional null-space corresponding to each $\lambda = \lambda_k$ for $k \geq 2$. An element of the null-space corresponding to $k \geq 2$, in polar components, is of form:

$$\begin{aligned} \boldsymbol{X}(r,\theta) &= \Big((v_r, v_\theta), 0, \big[N_{rr}, N_{\theta\theta}, N_{r\theta}\big], \big[0,0,0\big], (0,0), (0,0,s_3), (n_r, n_\theta, 0)\Big)^\top \\ &= A_k\Big(\big(\phi_r(r)\cos k\theta, \phi_\theta(r)\sin k\theta\big), 0, \big[\Phi_{rr}(r)\cos k\theta, \Phi_{\theta\theta}(r)\cos k\theta, \Phi_{r\theta}(r)\sin k\theta\big], \\ &\big[0,0,0\big], (0,0), \big(0,0,\phi_{s3}\sin k\theta\big), \big(\phi_{nr}\sin k\theta, \phi_{n\theta}\cos k\theta, 0\big)\Big)^\top + \\ &+ B_k\Big(\big(\psi_r(r)\sin k\theta, \psi_\theta(r)\cos k\theta\big), 0, \big[\Psi_{rr}(r)\sin k\theta, \Psi_{\theta\theta}(r)\sin k\theta, \Psi_{r\theta}(r)\cos k\theta\big], \\ &\big[0,0,0\big], (0,0), \big(0,0,\psi_{s3}\cos k\theta\big), \big(\psi_{nr}\cos k\theta, \psi_{n\theta}\sin k\theta, 0\big)\Big)^\top \\ &= \boldsymbol{X}_1(r,\theta) + \boldsymbol{X}_2(r,\theta), \end{aligned} \qquad (4.30)$$

where $\phi_r, \psi_r, \ldots$ are explicitly found but due their cumbersome expressions we do not include them in the main text in this article.[1]

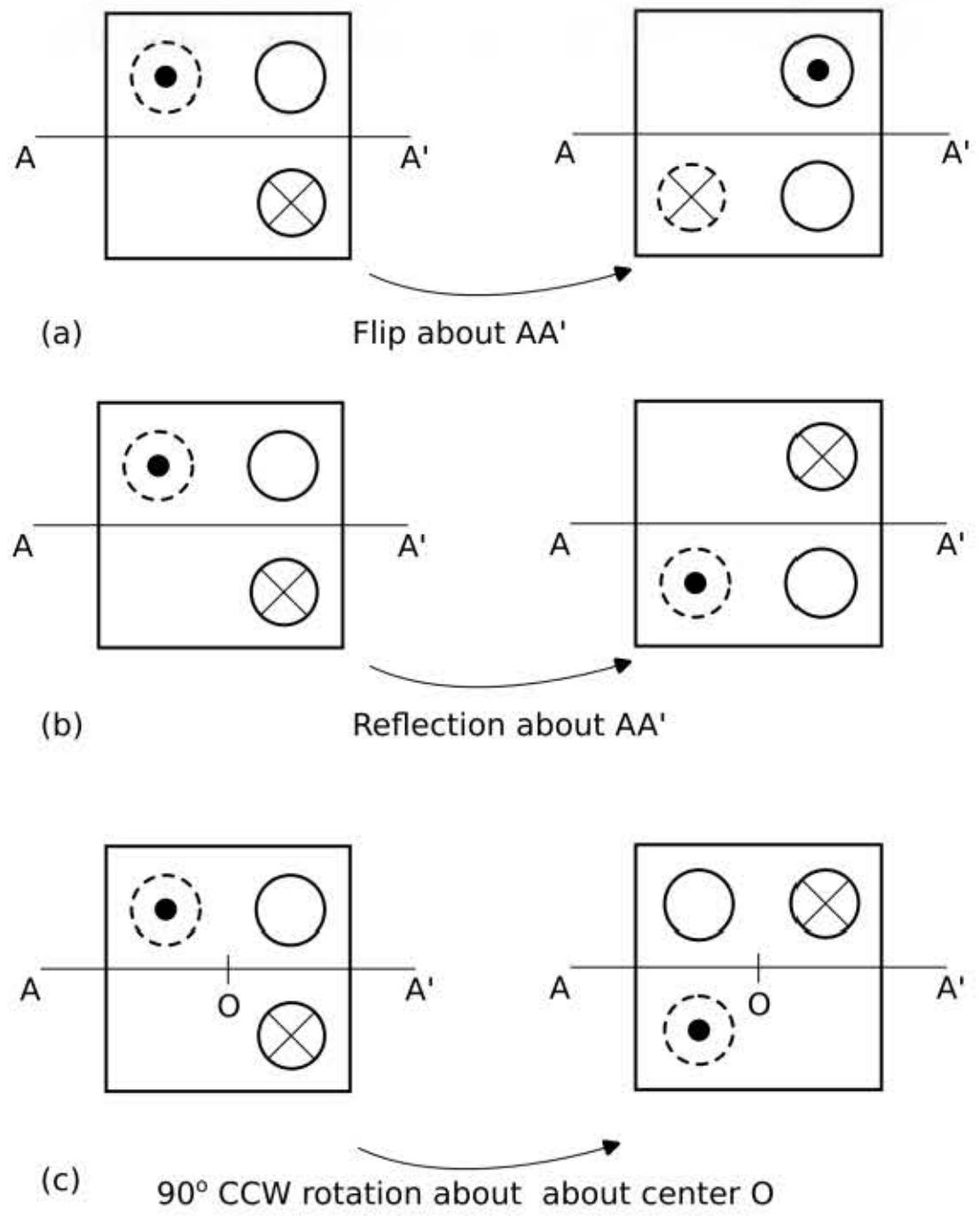

Figure 5: Operations of $\mathtt{Z}_4$ group. CCW denotes counter-clockwise. $\odot$ denotes tip of an arrow coming out of the plane and $\otimes$ denotes backside of the same arrow going into the plane.

Based on above expression (4.30) of an element of the null-space corresponding to $k \geq 2$, it is clear that we get a two-dimensional null space for each critical value $\lambda_k$. The null-space for $\lambda_k$ is expressed as linear combination of two null vectors corresponding to $A_k$ and $B_k$ in (4.30), denoted by $\boldsymbol{X}_1 \equiv \boldsymbol{X}_1(r, \theta)$ and $\boldsymbol{X}_2 \equiv \boldsymbol{X}_2(r, \theta)$ respectively. The null vector $\boldsymbol{X}_1$ is static under the action of the union of dihedral group of order $2k$, denoted as $\mathtt{D}_k$, and the flip (180 deg rotation) about the diameter of plate; the corresponding symmetry group is denoted by $\mathtt{Z}_k$. See the schematic illustration in Fig. 5 for the symmetry group $\mathtt{Z}_4$ (for $\boldsymbol{X}_1$); operations like those in Fig. 5(b) and Fig. 5(c) together form $\mathtt{D}_k$, while Fig. 5(a) shows the Flip operation. The symmetry group of $\boldsymbol{X}_2$ is obtained by conjugating $\mathtt{Z}_k$ by rotation of $\pi/2k$. See the schematic illustration Fig. 6 for the symmetry of $\boldsymbol{X}_2$ for $k = 4$; Fig. 6(a,b,c) correspond to operations (by rotation) conjugate to Fig. 5(a,b,c), respectively.

### *4.6. Out-of-plane perturbed solution relative to base solution*

Proceeding in the same way as in the previous case of in-plane perturbed solution relative to base solution, for example using the fact the displacement fields are periodic with respect to variable $\theta$, the out-of-plane displacement $z$ admits its representation as (complex) Fourier series in $\theta$,

$$z(r, \theta) = \sum_{k=-\infty}^{\infty} z_k(r) \exp(ik\theta). \tag{4.31}$$

[1] The detailed expressions for the components of $\boldsymbol{X}_1$ are, however, included in the supplementary S1.

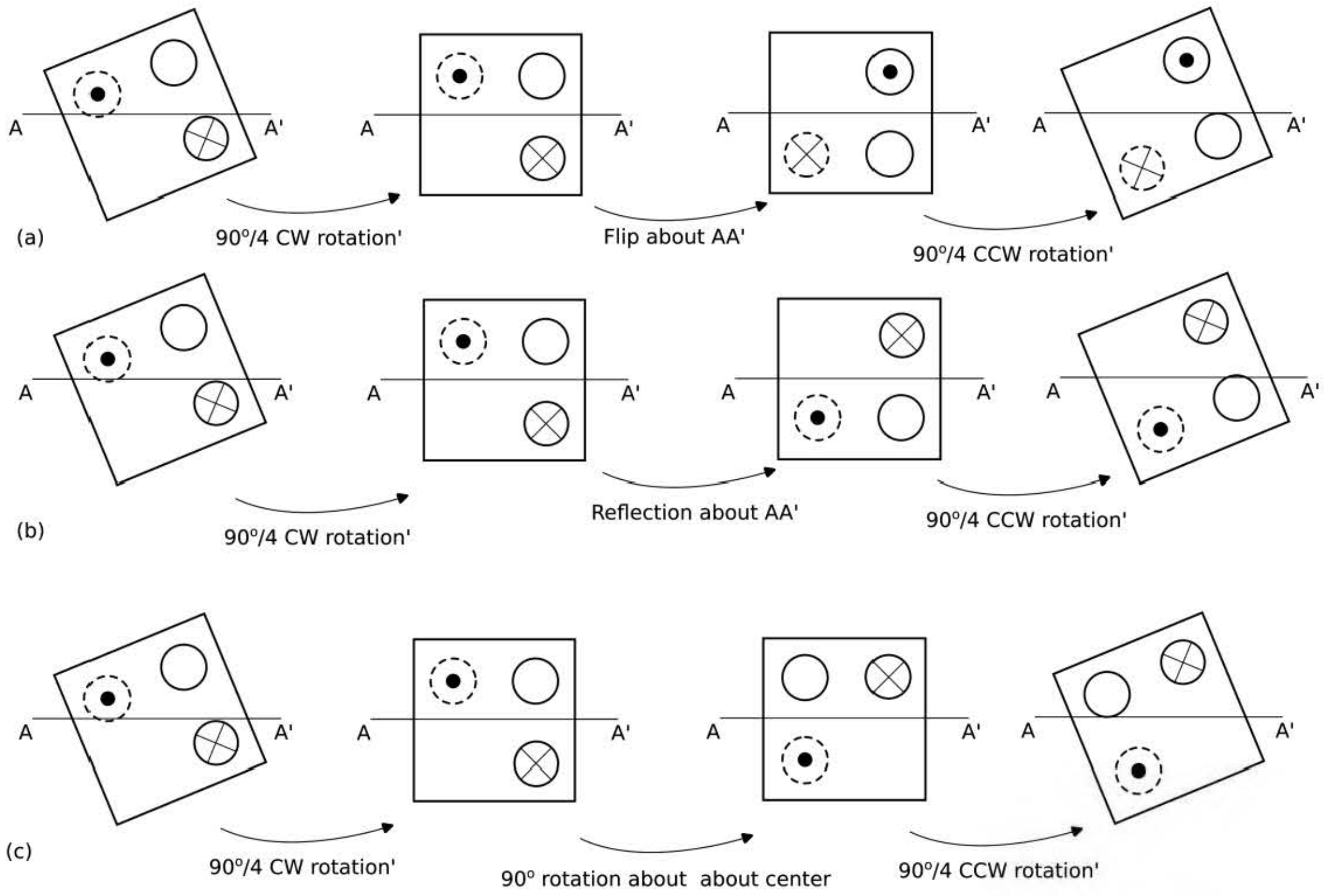

Figure 6: Operations of conjugate of $Z_4$ group. Other details same as in Fig. 5.

The expression (4.31) is substituted in (D.2) and the coefficient of $\exp(ik\theta)$ is collected to obtain a ordinary differential equation in $r$. The ordinary differential equation is solved to obtain $z_k(r)$ in (4.31). The general solution of (D.2), ignoring the terms which are singular at $r = 0$, is given as follows:

- For $0 < \lambda < 1$ (when the plate is stretched in intermediate configuration),

$$
\begin{aligned}
z(r,\theta) =& \sum_{k=-\infty}^{\infty} \Big( E_k r^{|k|} + F_k I_k(\sqrt{|a_\lambda|}r) \Big) \exp(ik\theta) \\
=& E_0 + \sum_{k=2}^{\infty} \Big( (A_k r^k + C_k I_k(\sqrt{|a_\lambda|}r)) \cos k\theta + \\
& + (B_k r^k + D_k I_k(\sqrt{|a_\lambda|}r)) \sin k\theta \Big);
\end{aligned}
\tag{4.32}
$$

- For $\lambda > 1$ (when the plate is compressed in intermediate configuration),

$$
\begin{aligned}
z(r,\theta) =& \sum_{k=-\infty}^{\infty} \Big( E_k r^{|k|} + F_k J_k(\sqrt{|a_\lambda|}r) \Big) \exp(ik\theta) \\
=& C_0 J_0(\sqrt{|a_\lambda|}r) + \sum_{k=1}^{\infty} \Big( (A_k r^k + C_k J_k(\sqrt{|a_\lambda|}r)) \cos k\theta + \\
& + (B_k r^k + D_k J_k(\sqrt{|a_\lambda|}r)) \sin k\theta \Big).
\end{aligned}
\tag{4.33}
$$

In (4.32) and (4.33), $J_k$ is the Bessel's function of first kind, $I_m$ is the modified Bessel's function of first kind and $a_\lambda = 12(1+\nu)(\lambda - \lambda^2)/h^2$.

The expressions (4.32) and (4.33) are substituted in boundary conditions (D.7) and (D.3d), and the system of equations is solved to get the critical values of $\lambda$ and corresponding non-trivial $E_k$ and $F_k$. Due to the presence of Bessel's function in the expressions (4.32) and (4.33), the critical values of $\lambda$ and corresponding $E_k, F_k$ are obtained numerically.

The following properties are observed regarding the critical values of the bifurcation parameter $\lambda$:

- As the boundary conditions only contain even derivatives with respect to $\theta$, $\lambda_k = \lambda_{-k}$ and $E_k : F_k = E_{-k} : (-1)^k F_{-k}$ as $|k| = |-k|$. In this case, we use the property of Bessel's functions, $(-1)^k I_k = I_{-k}$ and $(-1)^k J_k = J_{-k}$.
- For $0 < \lambda < 1$, a single $\lambda_k$ is found for each $k \geq 2$. For $k = 0$, the rigid displacement mode is found for all values of $\lambda$. There are no non-trivial solutions corresponding to $k = 1$.
- For $\lambda \in \mathbb{R}^+$, infinitely many critical points are found for each $k$. The $n^{th}$ critical value corresponding to $k$ is denoted by $\lambda_{k,n}$ and corresponding coefficients are denoted by $E_{k,n}$ and $F_{k,n}$.

The coefficient $E_0$ is arbitrary for all $\lambda$ as it gives the rigid motion in $\boldsymbol{e}_3$ direction. Hence, only $F_{0,n}$ is considered in the general solution corresponding to $k = 0$. For, $k = 0$ case, the null space is one dimensional and has form:

$$\begin{aligned}\boldsymbol{X}(r,\theta) &= \Big((0,0), y, \big[0,0,0\big], \big[M_{rr}, M_{\theta\theta}, 0\big], \big(f_r, 0\big), \big(0,0,0\big), \big(0,0,0\big)\Big)^\top \\ &= C\Big((0,0), \phi(r), \big[0,0,0\big], \big[\Phi_{rr}(r), \Phi_{\theta\theta}(r), 0\big], \big(\phi_{f_r}(r), 0\big), \big(0,0,0\big), \big(0,0,0\big)\Big)^\top,\end{aligned} \tag{4.34}$$

where $\phi, \Phi_{rr}, \ldots$ are explicitly found but due their cumbersome expressions we do not state in this article (see Footnote 1.)

Similar to the planar case, for each critical $\lambda_k$ for $k \geq 1$, there is a two-dimensional null-space of $\mathfrak{L}(\lambda_k)$. The elements of the null space can be expressed as:

$$\begin{aligned}\boldsymbol{X}(r,\theta) &= \Big((0,0), y, \big[0,0,0\big], \big[M_{rr}, M_{\theta\theta}, M_{r\theta}\big], \big(f_r, f_\theta\big), \big(s_r, s_\theta, 0\big), \big(0,0,n_3\big)\Big)^\top \\ &= A_k\Big((0,0), \phi(r)\cos k\theta, \big[0,0,0\big], \big[\Phi_{rr}(r)\cos k\theta, \Phi_{\theta\theta}(r)\cos k\theta, \Phi_{r\theta}(r)\sin k\theta\big], \\ &\quad \big(\phi_{fr}(r)\cos k\theta, \phi_{f\theta}(r)\sin k\theta\big), \big(\phi_{sr}(r)\sin k\theta, \phi_{s\theta}(r)\cos k\theta, 0\big), \big(0,0,\phi_{n3}\sin k\theta\big)\Big)^\top \\ &\quad + B_k\Big((0,0), \psi(r)\sin k\theta, \big[0,0,0\big], \big[\Psi_{rr}(r)\sin k\theta, \Psi_{\theta\theta}(r)\sin k\theta, \Psi_{r\theta}(r)\cos k\theta\big], \\ &\quad \big(\psi_{fr}(r)\sin k\theta, \psi_{f\theta}(r)\cos k\theta\big), \big(\psi_{sr}(r)\cos k\theta, \psi_{s\theta}(r)\sin k\theta, 0\big), \big(0,0,\psi_{n3}\cos k\theta\big)\Big)^\top \\ &= \boldsymbol{X}_1(r,\theta) + \boldsymbol{X}_2(r,\theta),\end{aligned} \tag{4.35}$$

where $\phi, \psi, \Phi_{rr}, \ldots$ are explicitly found; again, due their cumbersome expressions we do not include them in this main text. (see Footnote 1).

For each critical value $\lambda_k$, it is evident that we get a two-dimensional null space, that is the null-space for $\lambda_k$ is expressed as linear combination of two null vectors corresponding to $A_k$ and $B_k$ in (4.35), denoted by $\boldsymbol{X}_1 \equiv \boldsymbol{X}_1(r,\theta)$ and $\boldsymbol{X}_2 \equiv \boldsymbol{X}_2(r,\theta)$ respectively. One of the null vectors, that is $\boldsymbol{X}_1$, is static under actions of elements of Dihedral group of order $2k$, denoted as $\mathtt{D}_k$, while the other null vector, that is $\boldsymbol{X}_2$, is static under symmetry group given by the union of rotation group $\mathtt{C}_k$ and flip about the diameter. See the schematic illustration in Fig. 5(b,c) for the symmetry group $\mathtt{D}_4$ (for $\boldsymbol{X}_1$), while see the schematic illustration Fig. 5(a,c) for the symmetry of $\boldsymbol{X}_2$.

Up to this point in this article, we have formulated the problem of mechanics of the rod-plate system as a non-linear operator $\mathfrak{F}$ (4.10) with $(\mathbf{0}, \lambda)$ as its trivial solution. Then we proceeded with obtaining the linearization $\mathfrak{L}(\lambda)$ of the operator $\mathfrak{F}$ about the trivial solution branch in order to investigate the loss of uniqueness of the trivial solution branch with respect to the bifurcation parameter $\lambda$. Closed form and semi analytical expressions were found for critical values of $\lambda$ where the linear operator $\mathfrak{L}(\lambda)$ (4.21) becomes singular, implying a possibility of bifurcation from the trivial solution branch. The values of bifurcation parameter $\lambda$ and corresponding null spaces were determined in closed form for planar modes and semi analytically for out of plane modes.

As discussed in the introduction, structures similar to the rod-plate system occur in biological systems, thermal buckling problems and soft robotics. Therefore, in the next section, numerical evaluations of the critical values of the parameters are carried out by choosing four sets of structure parameters representing metals, biological materials and hypothetical combination of soft-elastic and stiff-elastic materials. To further understand the dependence of critical values of $\lambda$ on the structure parameters, parametric sweeps are performed around each parameter set. Finally, for validation, critical values of $\lambda$ and the mode shapes are compared to the ones obtained using finite element package, ABAQUS.

## 5. Illustrative numerical results

For a graphical illustration of the numerical evaluation of theoretical results, four sets of structure parameters are used, as described in the following:

A — In this case, the plate material is chosen as Copper and the the rod material is chosen as Steel. See column A in Table 1 and the caption.

(5.1)

B — The plate is assumed to be made of soft organic material (pumpkin) and the rod is assumed to be made of stiffer pumpkin skin ( (Gaibotti et al., 2024)). As the young modulus ratio is provided in the reference, an arbitrary unit 'E' is considered, which gets eliminated during the non-dimensionalization step. See column B in Table 1 and the caption.

(5.2)

C — A hypothetical configuration is considered in this case, where the plate's Young's modulus is two orders of magnitude lower than the Young's modulus of rod. See column C in Table 1 and the caption.

(5.3)

D — A hypothetical configuration is also considered in this case, where the plate's Young's modulus is two orders of magnitude higher than the Young's modulus of rod. See column D in Table 1 and the caption.

(5.4)

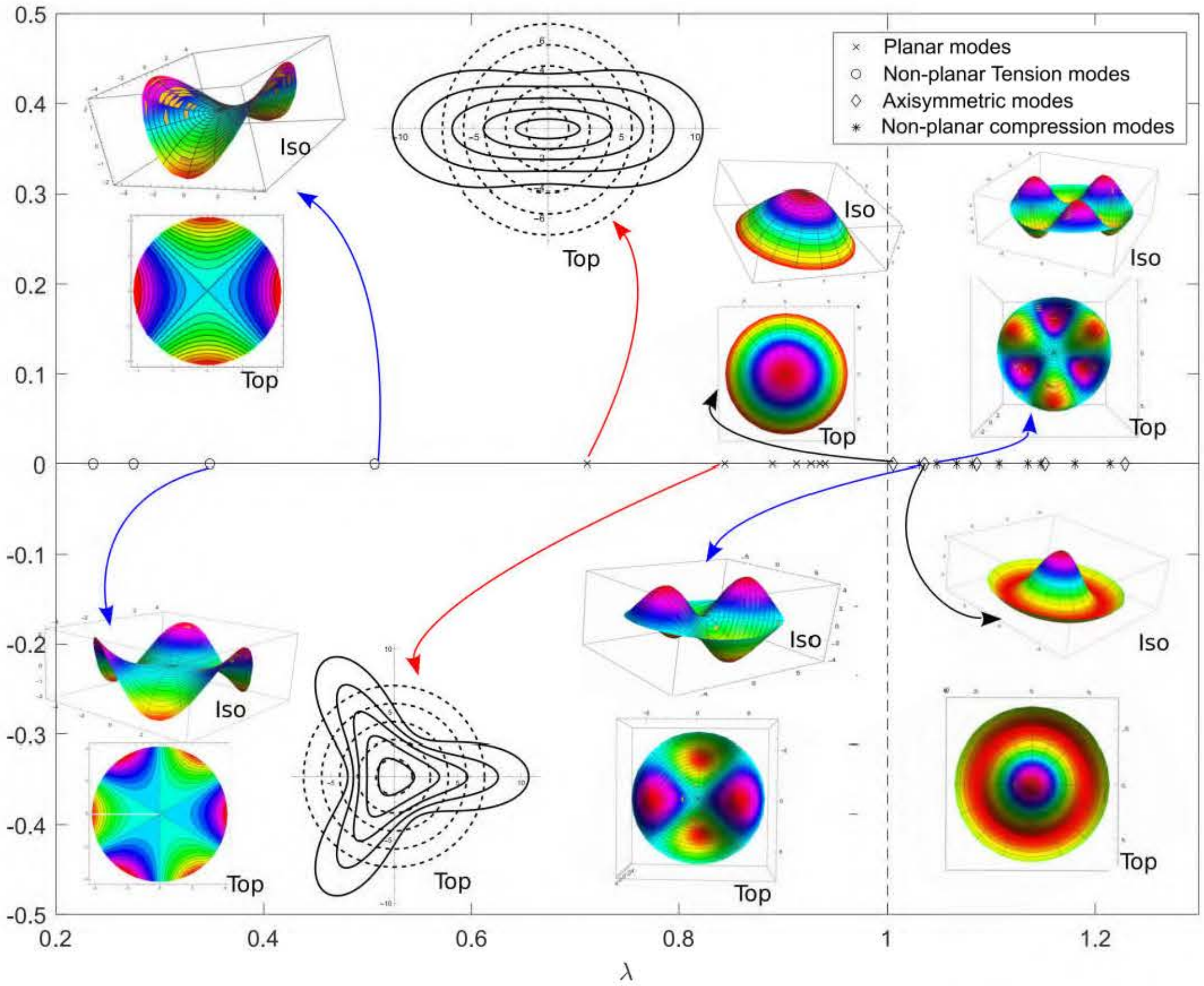

Figure 7: Selected critical values and corresponding mode-shapes for structure parameter set A (5.1) (see also Table 1, Table 2). 'Iso' stands for isometric view, 'Top' stands for top view of the mode-shapes. Uncolored plots of mode-shapes are inherently planar (shown total two such modes).

| Structure parameters | A | B | C | D |
|---|---|---|---|---|
| $E_{plate}$ | 110 GPa | 0.6 E | 1 GPa | 100 GPa |
| $\nu_{plate}$ | 0.34 | 0.5 | 0.3 | 0.3 |
| $E_{rod}$ | 200 GPa | 1 E | 100 GPa | 1 GPa |

Table 1: Physical (dimensional structure) parameter values considered for the analysis. In all cases, $\nu_{rod} = 0.3$ is chosen as the Poisson's ratio of the rod; $\mathfrak{D} = 0.028$m is the diameter of center-line of stress-free circular rod; $\mathtt{d}_r = h_{plate} = 0.002$m where $\mathtt{d}_r$ is the diameter of rod cross-section and $h_{plate}$ is the thickness of plate.

To calculate the bending stiffness and torsional stiffness of the rod from the physical (structure) parameters of Table 1, the rod cross-section is assumed to be circular and following (Hafner and Bickel, 2023), the formulae for bending stiffness and twisting stiffness that have been used are:

$$\beta = E_{rod} I_{rod}, \quad \gamma = \frac{E_{rod}}{2(1+\nu_{rod})} J_{rod}, \tag{5.5}$$

where $I_{rod}$ $(= \pi \mathtt{d}_r^4/64)$ and $J_{rod}$ $(= \pi \mathtt{d}_r^4/32)$ are the area moment of inertia and polar moment of inertia of the circular cross-section of rod (of diameter $\mathtt{d}_r$).

The non-dimensional structure parameters, as listed in Table 2, are determined using the scheme presented in (4.13) in §4.2, with length scale $L = \mathfrak{D}/2$. Due to the choice of length scale, the parameter $\alpha = 1$ for all structure parameter sets (5.1)–(5.4). Note that the thickness of the plate and the diameter of cross-section of the rod are assumed to be the same (see caption of Table 1).

| Parameter values | A | B | C | D |
|---|---|---|---|---|
| $\beta$ | 0.00276149 | 0.00214668 | 0.156278 | 0.0000156278 |
| $\gamma$ | 0.00212422 | 0.00165129 | 0.120214 | 0.0000120214 |
| $\nu$ | 0.34 | 0.5 | 0.3 | 0.3 |

Table 2: Values corresponding to the chosen (dimensionless) structure parameter sets listed as (5.1)–(5.4) (see also Table 1). In all cases $h = 0.142857$.

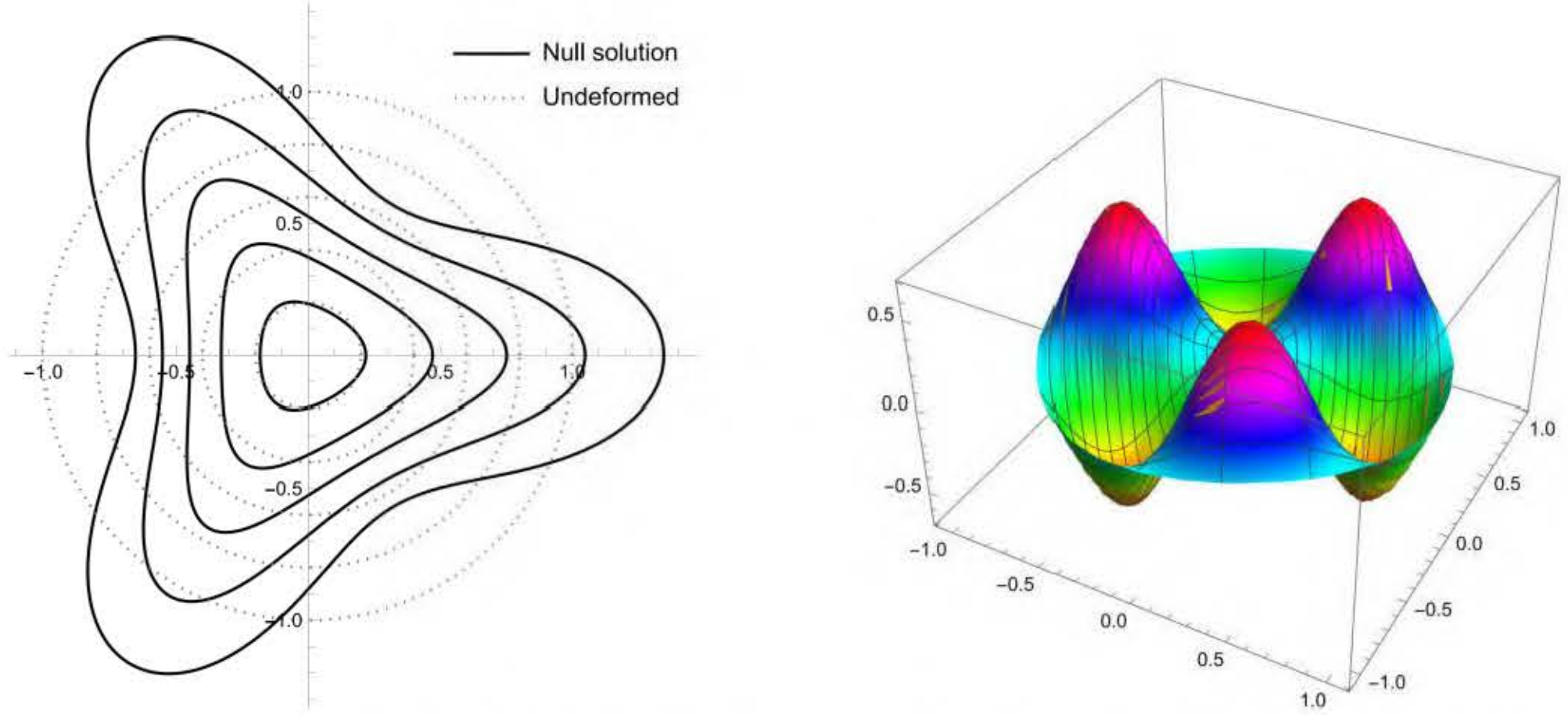

(a) Mode-shape satisfying $\mathtt{Z}_3$ symmetry. See also the same mode-shape shown in Fig. 7.

(b) Mode-shape satisfying $\mathtt{D}_3$ symmetry.

Figure 8: Illustrative mode-shapes in the symmetry groups $\mathtt{Z}_3$ and $\mathtt{D}_3$.

Fig. 7 shows an overview of all the critical points and corresponding mode shapes corresponding to structure parameter set A (5.1) from Table 2. The horizontal axis represents $\lambda$ and the critical values of $\lambda$ are also indicated by several different symbols. As can be seen in Fig. 7, the null spaces have distinct symmetries. In Fig. 7, the symbol $\circ$ represents the critical points corresponding to $\mathtt{D}_k$ symmetric modes (non-planar), while the symbol $\times$ represents the critical points corresponding to $\mathtt{Z}_k$ symmetric modes (planar), both in case of $\lambda < 1$, i.e., when the plate is in tension; the symbol $\diamond$ represents the critical points corresponding to axisymmetric modes (non-planar), while, the symbol $*$ represents the critical points corresponding to $\mathtt{D}_k$ symmetric modes (non-planar), both in case of $\lambda > 1$, i.e., when the plate is in compression.

Few more comments are in order for Fig. 7. With the symbol $\circ$, only the first four critical values of $\lambda$, discussed in §4.6, that for $k = 2, 3, 4, 5$; note that these critical values are obtained numerical as there is no closed form expression and the other critical values corresponding to higher $k$ are not shown though they lie in the same interval $(0, 1)$. Note that there are only a finite number of critical points of $\lambda$ for such $\mathtt{D}_k$ symmetric modes when the plate is in tension ($\lambda < 1$); we present a convincing argument later in the discussion of parameter sweeps. With the symbol $\times$, only the first seven critical values of $\lambda$, that with $k = 2, \dots, 8$, are shown using the closed form expression (4.28) and the other critical values corresponding to higher $k$ (but with total number finite) are not shown though they are not found to exhibit any monotonic behavior towards $\lambda = 1$ (see Remark 6). With the symbol $\diamond$, that is the axi-symmetric case, only those critical values of $\lambda$ are shown such that $\lambda \in (1, 1.2)$ while there are certainly an infinite number of critical values for $\lambda > 1$. With the symbol $*$, the critical values of $\lambda$ corresponding to $k = 2, 3, 4$ such that $\lambda \in (1, 1.2)$ are shown; however, in each case of $k$ there are also an infinite number of critical values and these are not shown.

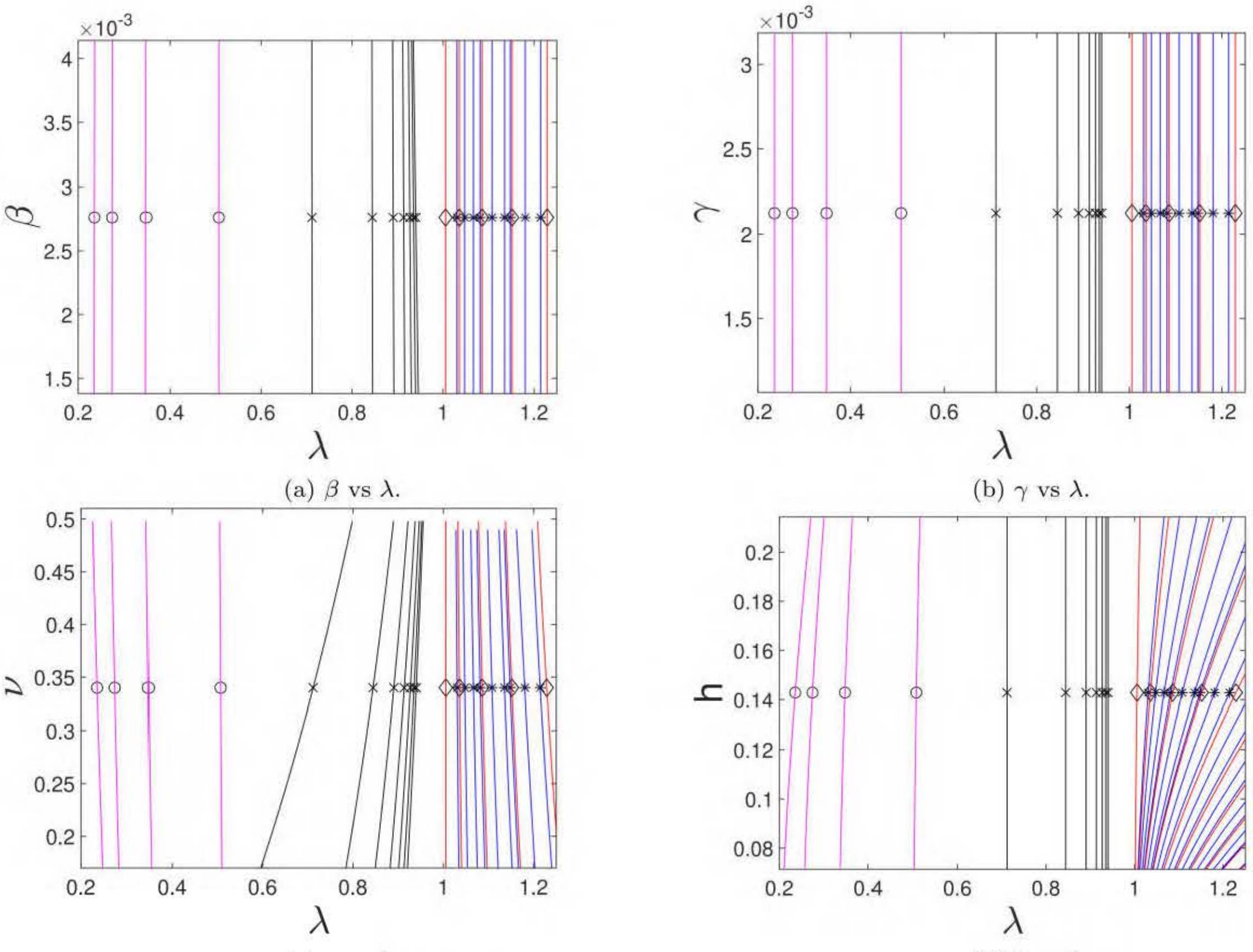

(a) $\beta$ vs $\lambda$.

(b) $\gamma$ vs $\lambda$.

(c) ν vs $\lambda$.

(d) $h$ vs $\lambda$.

Figure 9: Parameter sweeps around structure parameter set A (5.1) (see also Table 2).

Also in Fig. 7, we have plotted the null solution $\boldsymbol{X}_1$ for planar case where we used Eq. (4.30) (see also the paragraph following it), for axisymmetric case we used Eq. (4.34), and for non-planar, non-axisymmetric case we have plotted $\boldsymbol{X}_1$ using Eq. (4.35) (see also the paragraph following it).

Out of the eight cases presented in Fig. 7, we have selected the one on top-right and bottom-left to present separately to highlight the features more clearly, specially regarding the symmetry of modes. The null solutions shown in Fig. 8a and Fig. 8b correspond to $\mathtt{Z}_3$ and $\mathtt{D}_3$ symmetries, respectively.

Parameter sweep study is conducted for critical points for a span of around 50% change in each parameter value for each of the four choices of structure parameters (5.1)–(5.4) (see also the details of numerical values of structure parameters given in Table 2, with physical values in Table 1), and the resulting plots are provided in Fig. 9–Fig. 12. In these figures, the symbol $\circ$ represent the critical points corresponding to $\mathtt{D}_k$ symmetric modes for $\lambda < 1$, i.e., the plate is in tension, the symbol $\times$ represent the critical points corresponding to $\mathtt{Z}_k$ symmetric modes for $\lambda < 1$, i.e., the plate is in tension and planar, the symbol $\diamond$ represent the critical points corresponding to axisymmetric modes for $\lambda > 1$, i.e.,the plate is in compression, the symbol $*$ represent the critical points corresponding to $\mathtt{D}_k$ symmetric modes for $\lambda > 1$, i.e., the plate is in compression. The curves colored in Magenta are parametric sweeps of $\mathtt{D}_k$ symmetric modes ($\lambda < 1$) when the plate is in tension, black colored curves are of $\mathtt{Z}_k$ symmetric modes, red colored curves are for axisymmetric modes and blue colored curves are for $\mathtt{D}_k$ symmetric modes ($\lambda > 1$) when the plate is in compression.

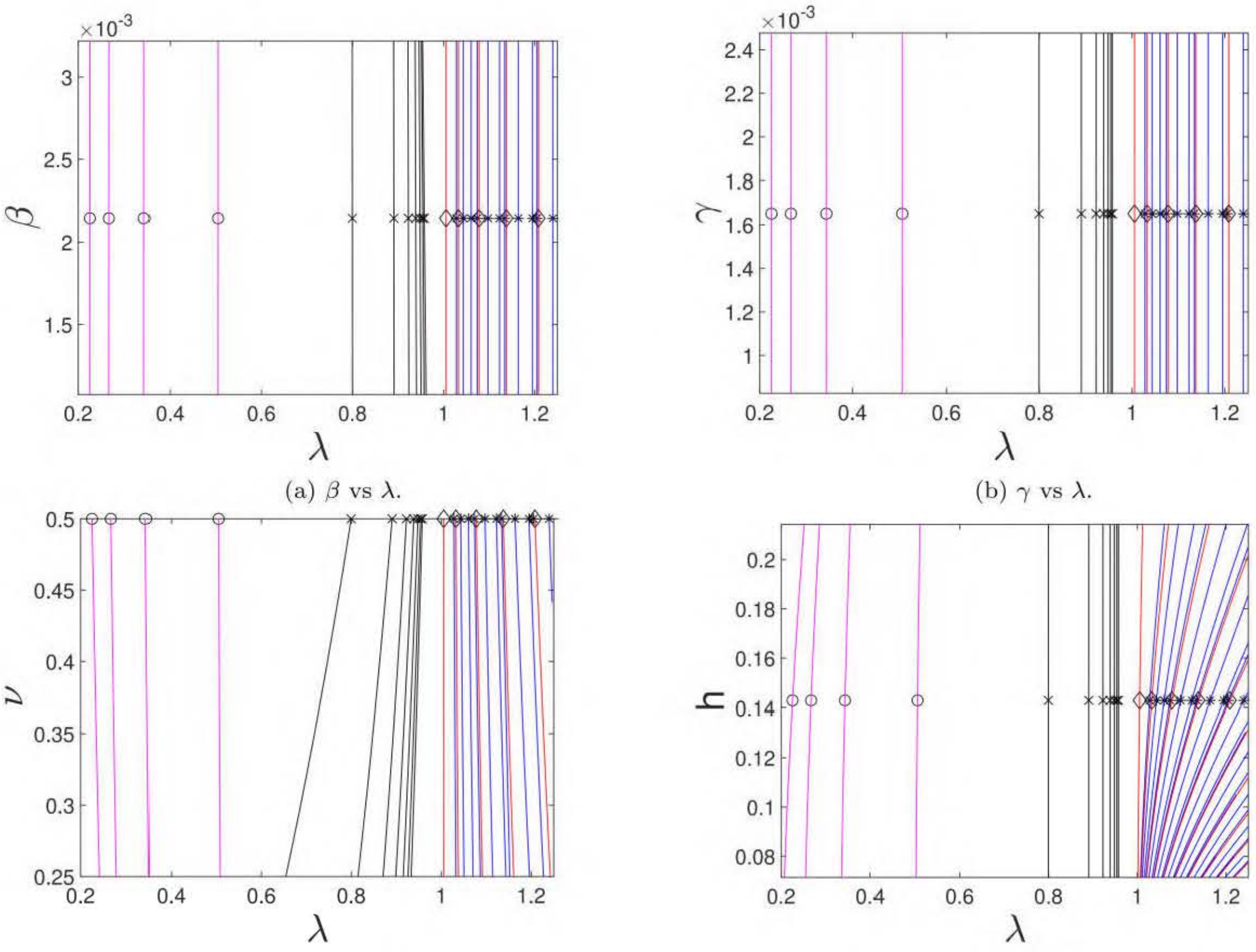

(a) $\beta$ vs $\lambda$.

(b) $\gamma$ vs $\lambda$.

(c) $\nu$ vs $\lambda$.

(d) $h$ vs $\lambda$.

Figure 10: Parameter sweeps around structure parameter set B (5.2) (see also Table 2).

Some notable trends are observed in the parametric sweep as described in the next few lines.

- The rod parameters have smaller impact on the critical value of the bifurcation parameter except when the rod material is considerably more stiff than the plate material, as in the case with structure parameter set C (5.3) (see also Table 1, Table 2) in Fig. 11.

- In the case of structure parameter C (5.3), the bending stiffness $\beta$ has no effect on the critical values of $\lambda$ for $\lambda > 1$. However, the bending stiffness $\beta$ has an impact on the critical values of $\lambda$ for $\lambda < 1$ region, i.e., for modes in which the plate is in tension.

- In Fig. 11(a), it is seen that for both the planar critical $\lambda$ sweeps (black curves) and non-planar critical $\lambda$ sweeps (magenta curve), some of the curves do not exist for high enough $\beta$, this is due to the fact that the rod becomes too stiff for the plate to cause the buckling of rod. Note that, with the plate model assumed, inspection of (4.5) reveals that the stress in the homogeneously deformed plate is hydrostatic with the value $12\mathfrak{C}(1 + \nu)(1 - \lambda)$ so that as $\lambda \to 0+$, the hydrostatic stress in trivial configuration remains finite. This implies that there are only finite number of critical points of $\lambda$ for both $\mathtt{D}_k$ symmetric modes and $\mathtt{Z}_k$ symmetric modes when the plate is in tension ($\lambda < 1$).

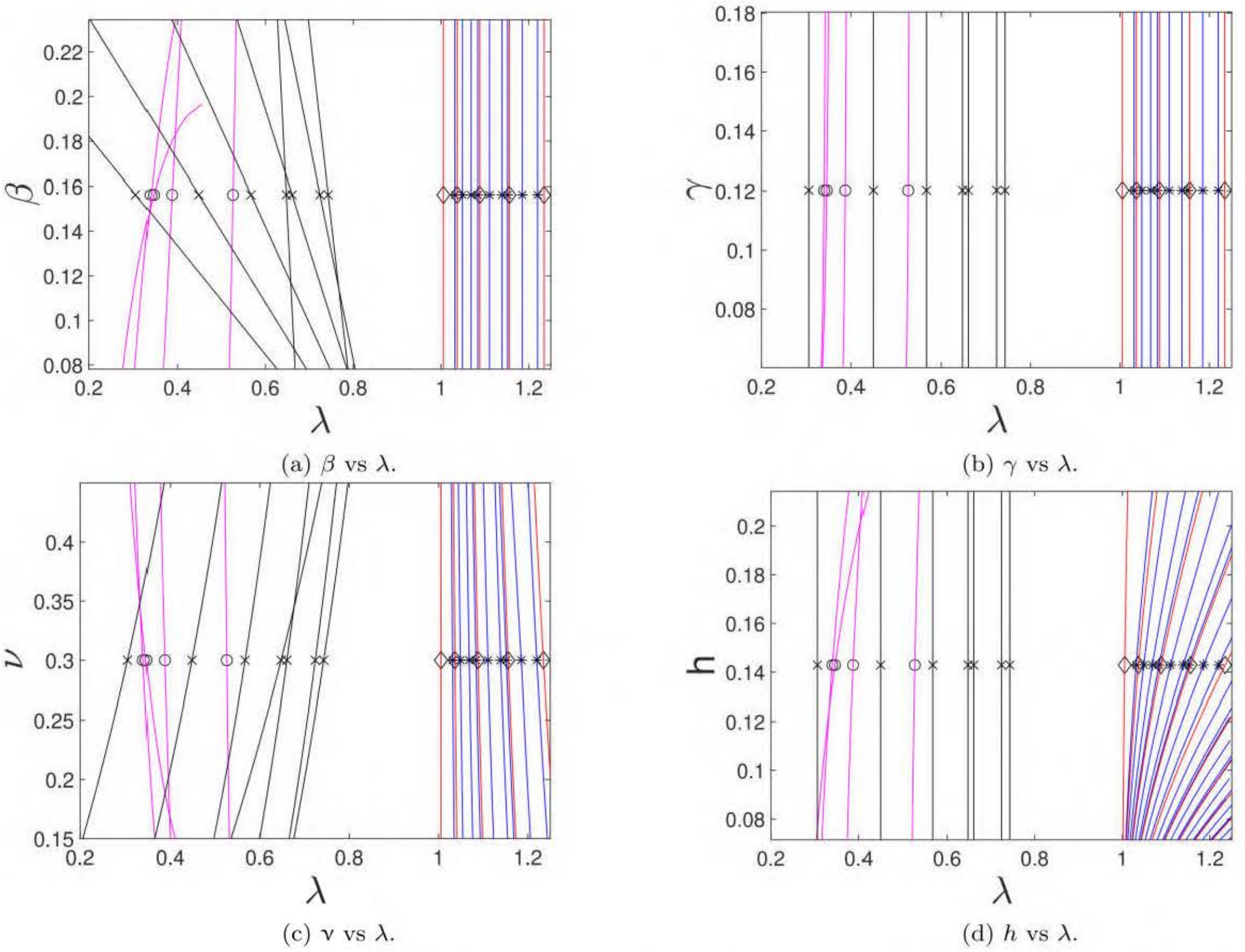

Figure 11: Parameter sweeps around structure parameter set C (5.3) (see also Table 2).

- The sweep of twisting stiffness $\gamma$ has no effect for structure parameter sets A, B and D and negligible effect on critical values corresponding to non-planar modes in $\lambda < 1$ region. This is because all the $\lambda < 1$ mode shapes are bending dominant and those of $\lambda > 1$ are rod independent.

- The Poisson's ratio $\nu$ appears in the expressions for plate membrane energy and the plate bending energy (1.7). Therefore it has an appreciable impact on critical values of $\lambda$ for parameter sweeps about all four structure parameter sets. It's effect is quite pronounced for the values corresponding to planar critical modes and is the highest, among all four choices, when the plate is considerably softer, i.e., for example the case of structure parameter set C (5.3) (see also Table 1, Table 2).

- The thickness of the plate $h$ appears only in the bending energy of the plate (1.7) relative to the membrane energy of the plate, therefore it has no effect on the planar critical modes. Its effect is observed the most prominently, among all four choices of structure parameters, only for the sweep around structure parameter set C (5.3).

- The critical values of $\lambda$, when $\lambda < 1$, also embody a peculiar behavior for structure parameter set C (5.3), that is when the buckling modes are dominated by rod buckling. It has been found that the parameter sweep lines cross each other, implying that for some parameter values of $\beta$, $\nu$ and $h$, the null spaces have dimensions higher than two.

An independent check on the critical points and instability modes has been also carried out in the context of the research problem analyzed in this article. For this independent check via FEM analysis using ABAQUS, the parameters corresponding to structure parameter set A (5.1), from Table 1, are used. The cross-section of the rod is assumed to be circular in order to preserve isotropic response of the rod. General purpose shell element is used for the plate and wire element is used to model the rod. A mesh size of 0.2 mm is used to mesh both rod and the plate. The linear buckling is carried out with respect to a uniform temperature load with constant expansion coefficient of 1% per °C. The thermal load is then converted to corresponding diameter ratio $\lambda$. The comparison of critical values computed from the finite element method and semi-analytical method for $\mathtt{D}_3$ symmetric mode shape is given in Table 3. The qualitative nature of buckled mode is obtained as shown in Fig 13c and Fig. 13d. However, the attempts for other structure parameter sets have not been successful as ABAQUS/Explicit uses general purpose shell elements which don't approximate to von-Kármán plate under arbitrary structural parameters, hence the critical points only match for thin plate and small buckling mismatch. The results shown in Table 3 only coincide for small deformations and the mode shapes compare well qualitatively with the theoretical results.

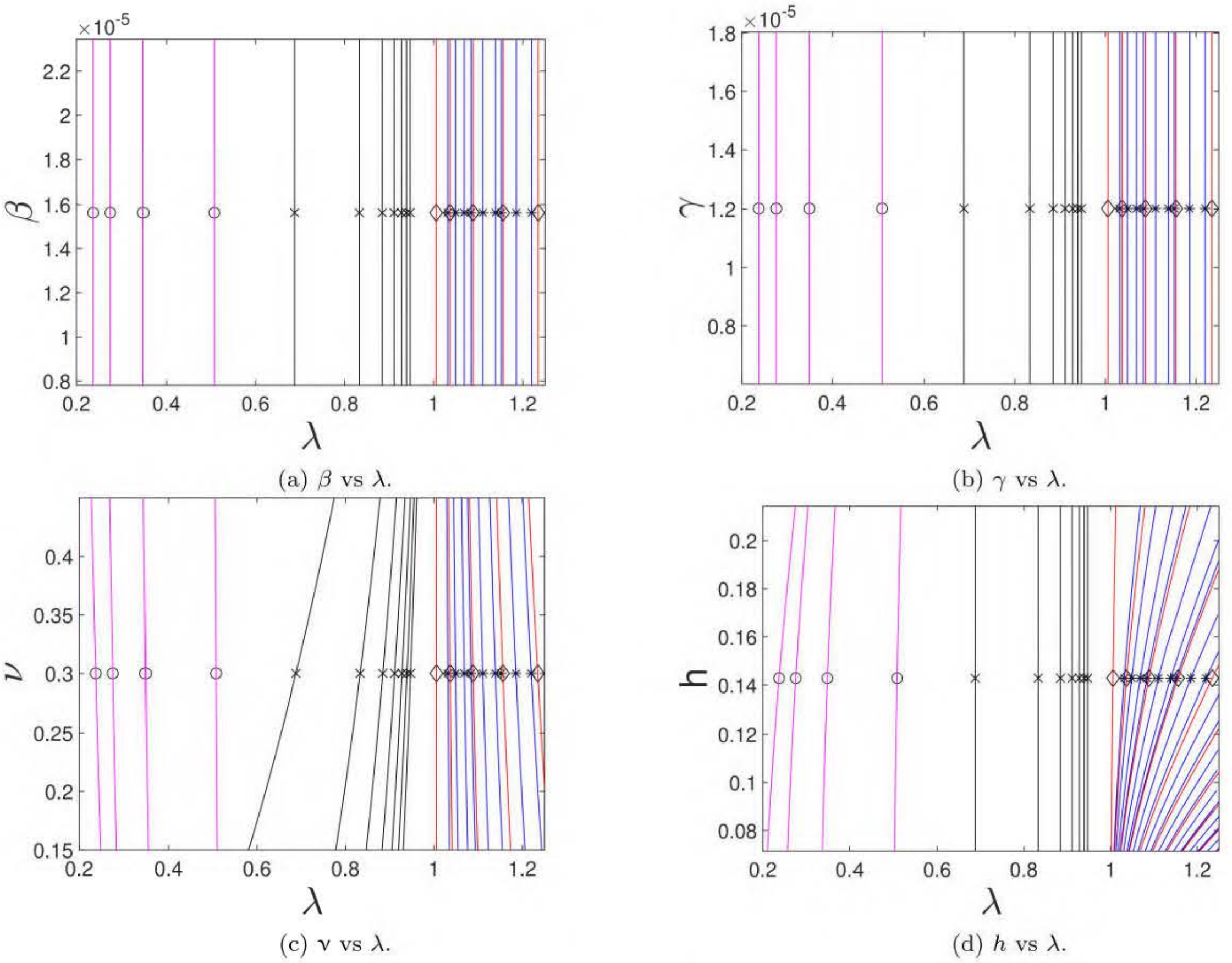

(a) $\beta$ vs $\lambda$.

(b) $\gamma$ vs $\lambda$.

(c) $\nu$ vs $\lambda$.

(d) $h$ vs $\lambda$.

Figure 12: Parameter sweeps around structure parameter set D (5.4) (see also Table 2).

| $\lambda$ Theoretical | 1.04746 | 1.10757 | 1.18063 | 1.26311 |
|---|---|---|---|---|
| $\lambda$ FEM | 1.0486 | 1.1236 | 1.2357 | 1.3125 |

Table 3: Critical diameter ratios for $\mathtt{D}_3$ symmetric modes.

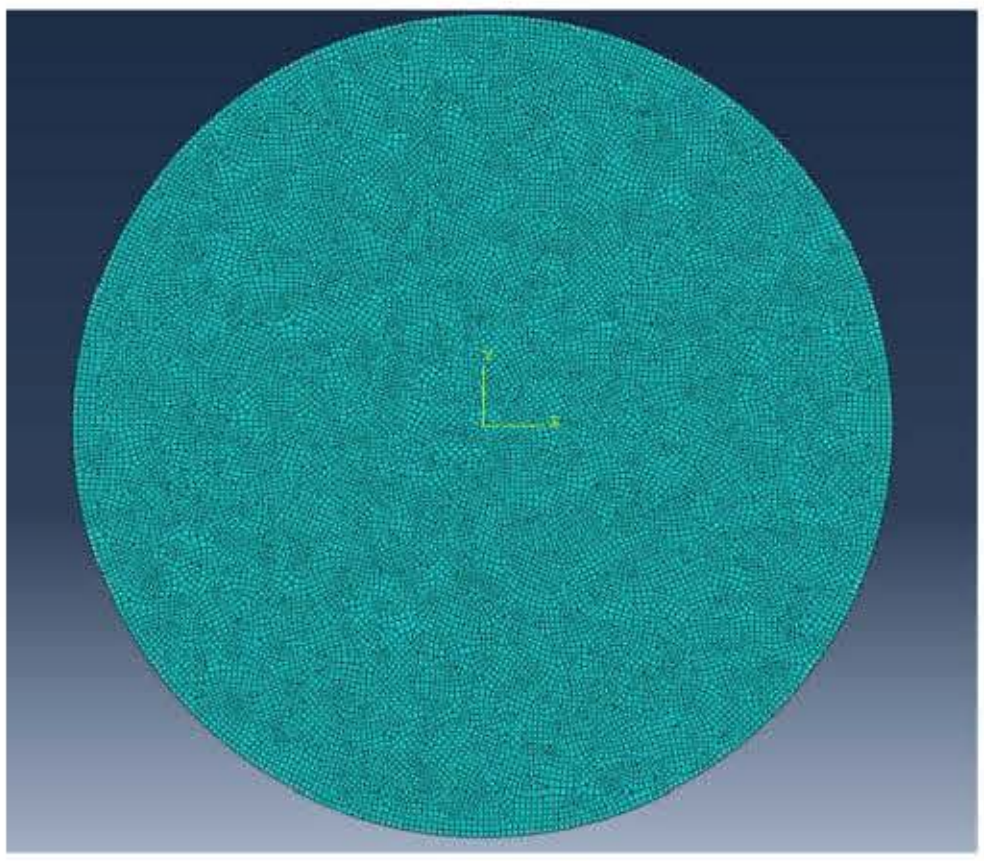

(a) Finite element mesh of the system in ABAQUS.

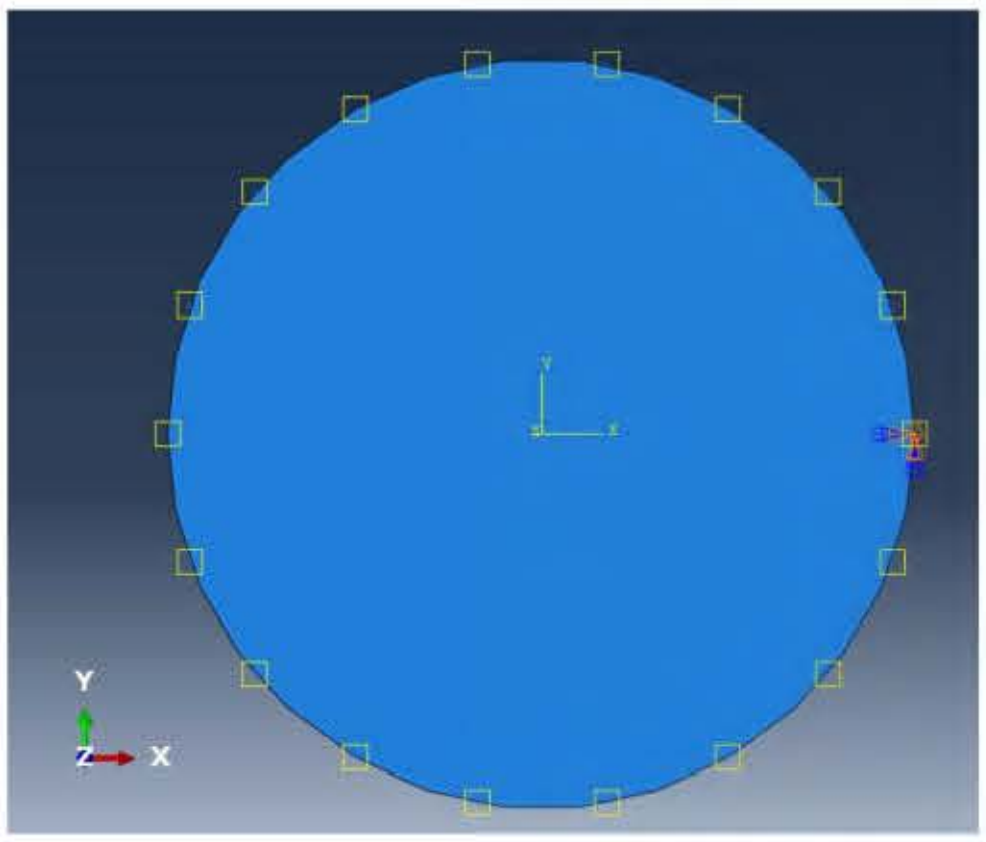

(b) Boundary conditions enforced in the FEM model using ABAQUS.

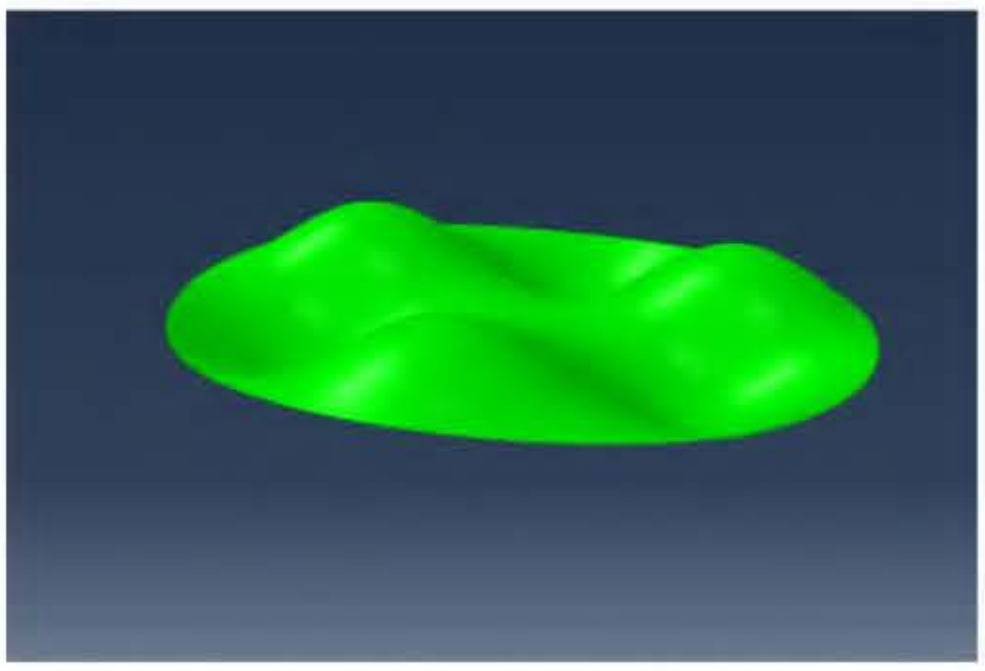

(c) Compressed-plate mode-shape with $\mathtt{D}_3$ symmetry obtained using ABAQUS.

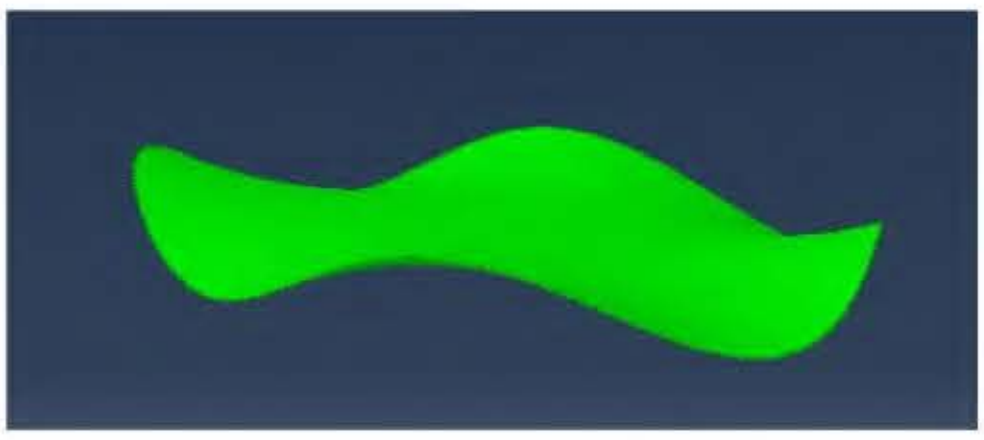

(d) Tension-in-plate mode-shape with $\mathtt{D}_4$ symmetry obtained using ABAQUS.

Figure 13: Illustrative FEM model based solution.

## 6. Discussion and concluding remarks

In this article, using the standard tools of calculus of variations and nonlinear theory of elasticity, we have formulated the problem of equilibria of a rod-plate system. A mismatch of dimension between the rod and the plate boundary in their respective stress-free configurations, and the process of gluing of the inextensible circular rod to the edge of the extensible plate, causes the rod-plate system to develop internal stress in the natural planar configuration. We have formulated this problem as a non-linear operator with a flat state as its trivial solution for all physically reasonable values of the mismatch (the bifurcation parameter). The linearization of the nonlinear operator about the trivial solution branch is obtained in order to investigate the loss of uniqueness of the trivial solution branch with respect to the bifurcation parameter. Theoretical results have been found for the critical values of the bifurcation parameter, and corresponding modeshapes, in order to satisfy the necessary conditions for a bifurcation from the trivial solution branch. The critical values of bifurcation parameter and corresponding null spaces are provided in closed form for in-plane modes and in semi-analytical form for out-of-plane modes.

The rod-plate system like structures occur in biological systems, in problems involving buckling due to thermal effects, and in applications of actuation in soft robotics, for example. Therefore, for numerical illustations four sets of structure parameters are chosen representing metals, biological materials and a hypothetical combination of soft-elastic and stiff-elastic materials. The behaviour of critical values of the bifurcation parameter in a neighbourhood of these four sets of structure parameters is also studied by performing parametric sweeps around each parameter set. Last but not the least, the critical values of bifurcation parameter and the mode shapes are also validated against the finite element package ABAQUS.

Several features of the problem are noteworthy as concluded from the analysis carried out in this work. A summary of some of these features is as follows.

From an overview of all the critical points and corresponding mode-shapes, we find that the null spaces have distinct symmetries (which has been exploited for post buckling analysis in our second research paper on this problem, as the symmetry can be used to reduce null space to one dimension). When the perimeter of stress-free plate is larger than the length of rod, i.e., $\lambda > 1$, the rod plays a passive role in buckled mode-shapes. This can be attributed to the inextensibility constraint on the rod, which make the rod-plate system behave like a plate with Dirichlet boundary. Thus, when $\lambda > 1$, there is similarity of the results with the well known results in plate buckling. In particular, there are infinite number of modes for any given set of structure parameters and also there are infinite number of critical points for each symmetry class of null solutions. For the case of $\lambda < 1$, there are a finite number of modes for any given set of structure parameters even for non planar modes.

From the the study involving sweep in structure parameters, it has been observed that the twisting stiffness has no impact on planar critical points with $\lambda < 1$, and non-planar modes with $\lambda > 1$, while extremely small effect on those of non-planar modes with $\lambda < 1$. This is expected because first two types don't have any twist in the rod, while the non-planar modes with plate in tension have small twist of the rod. When the plate's Young's modulus is considerably less than that of rod's material (the structure parameter set C (5.3), for example), we observe peculiar characteristics such that the fact that the parameter sweep curves of critical values of $\lambda$ cross for some values of structure parameters. This indirectly implies that the null spaces corresponding to these crossing points have dimensions higher than two, further necessitating more careful approach.

There are also some generalizations where the analysis of the article can still be applied. These generalizations correspond to relaxing some of the restrictions imposed in the modelling of the problem. For example, a model of plate that involves radial gradient of material properties, but otherwise remains isotropic and elastic, leads to a system for which the methods presented in this article can be applied and some of the conclusions may continue to hold. The rod model can be improved by incorporating chirality (Healey, 2002) and further the rod-plate junction can be extended to include non-hinge supports where the rod directors interact with plate tangent and normal. The assumed rod is circular in its stress-free reference configuration, however the same assumption can be relaxed by including a reference curvature suitably (Sharma et al., 2023) while the same framework can be retained.

There are some critical limitations to the current modelling of the problem.

While carrying out the simplification of the equilibrium equations in sections, § 4.2 and § Appendix D.2, the structure parameters $\beta$ and $\gamma$ are assumed to be non zero, therefore the results can't be applied for strings directly but via a careful analysis of possibly singular limit. Due to limitation of the chosen plate model, the results are not applicable for large deformations even if the rod model allows it. Note that the von-Kármán plate assumes non-linear stretch and linear curvature, therefore it cannot be applied for cases where the edge, being an independent structure of elastic material, has large deformations. The Kirchhoff rod assumption, provides high level of non-linearity but limits meaningful rod-plate interaction to $\lambda < 0$.

The symmetries of the null spaces can be exploited to study the post buckling behaviour of the buckling branch if it exists. The symmetries of the null-spaces can be used to construct a robust post-buckling analysis both numerically (Wohlever, 1993) and semi-analytically (Healey, 1988) which is currently being written as part of a separate exposition. Moreoever, different shell and rod models can be used to observe more interesting behaviour. In particular, the model can be expanded to accommodate larger deformations by using shell models instead of plate models.

### Acknowledgments

The authors thank Sanjay Dharmavaram for discussions during his visit to IIT Kanpur in December 2021.

### Appendix A. First variation of total energy

In its state of equilibrium, the mechanical system (composed of a plate with a boundary stiffened by rod) achieves an extremum of the energy functional (2.4). A necessary condition for this to happen is that the first variation of $\mathcal{V}$ becomes zero for all admissible variations of the relevant fields (Gelfand et al., 2000; Giaquinta and Hildebrandt, 2004). The variation of $\widetilde{\boldsymbol{v}}$ and $\widetilde{w}$ is denoted by

$$\widetilde{\delta\boldsymbol{v}}:\widetilde{\Omega}\to\operatorname{span}(\boldsymbol{e}_1,\boldsymbol{e}_2)\quad\text{and}\quad\widetilde{\delta w}:\widetilde{\Omega}\to\mathbb{R},\tag{A.1}$$

respectively. The variation of $\widetilde{\mathtt{R}}$ is expressed in terms of $\delta\widetilde{\boldsymbol{\theta}}:\widetilde{\Gamma}\to\mathbb{R}^3$ as

$$\delta\widetilde{\mathtt{R}}=\operatorname{skw}(\delta\widetilde{\boldsymbol{\theta}})\widetilde{\mathtt{R}}.\tag{A.2}$$

The variations of $\widetilde{\boldsymbol{r}}$, $n_{||}$, $\boldsymbol{n}^L$, and $\Upsilon$ are denoted by

$$\delta\widetilde{\boldsymbol{r}}:\widetilde{\Gamma}\to\mathbb{R}^3,\delta n_{||}:\widetilde{\Gamma}\to\mathbb{R},\delta\boldsymbol{n}^L:\widetilde{\Gamma}\to\mathbb{R}^3,\quad\text{and}\quad\delta\Upsilon:\widetilde{\Gamma}\to\mathbb{R},\tag{A.3}$$

respectively.

In order to determine the Euler Lagrange equations, it is assumed that (using componentwise qualification for vector and tensor fields)

- $\widetilde{\boldsymbol{v}}\in(\mathcal{C}^2_{\widetilde{\Omega}})^2$, $\widetilde{w}\in(\mathcal{C}^4_{\widetilde{\Omega}})$,
- $\widetilde{\boldsymbol{r}}\in(\mathcal{P}^2_{\widetilde{\Gamma}})^3$, $\widetilde{\mathtt{R}}\in\{\mathbf{T}\in(\mathcal{P}^2_{\widetilde{\Gamma}})^9:\mathbf{T}(\widetilde{\boldsymbol{x}})\in\mathrm{SO}(3)\}$,
- $n_{||}\in(\mathcal{P}^1_{\widetilde{\Gamma}})$, $\widetilde{\boldsymbol{n}}^L\in(\mathcal{P}^1_{\widetilde{\Gamma}})^3$, and $\Upsilon\in(\mathcal{P}^0_{\widetilde{\Gamma}})$.

The first variation of the potential energy $\mathcal{V}_P$ of plate (2.5) is found to be

$$\begin{aligned}\mathfrak{C}^{-1}\delta\mathcal{V}_P=&-\int_{\widetilde{\Omega}}\widetilde{\delta\boldsymbol{v}}\cdot(\widetilde{\nabla}\cdot\widetilde{\mathbf{N}})\mathrm{d}\widetilde{A}-\int_{\widetilde{\Omega}}\widetilde{\delta w}(\widetilde{\nabla}\cdot(\widetilde{\mathbf{N}}\widetilde{\nabla}\widetilde{w}+\widetilde{\nabla}\cdot\widetilde{\mathbf{M}}))\mathrm{d}\widetilde{A}+\\&+\int_{\partial\widetilde{\Omega}}\widetilde{\delta\boldsymbol{v}}\cdot(\widetilde{\mathbf{N}}\widetilde{\boldsymbol{l}})\mathrm{d}\widetilde{l}+\int_{\partial\widetilde{\Omega}}\widetilde{\delta w}\big((\widetilde{\mathbf{N}}\widetilde{\nabla}\widetilde{w}+\widetilde{\nabla}\cdot\widetilde{\mathbf{M}})\cdot\widetilde{\boldsymbol{l}}+\widetilde{\nabla}(\widetilde{\boldsymbol{t}}\cdot\widetilde{\mathbf{M}}\widetilde{\boldsymbol{l}})\cdot\widetilde{\boldsymbol{t}}\big)\mathrm{d}\widetilde{l}-\\&-\int_{\partial\widetilde{\Omega}}(\widetilde{\boldsymbol{l}}\cdot\widetilde{\mathbf{M}}\widetilde{\boldsymbol{l}})(\widetilde{\nabla}\widetilde{\delta w}\cdot\widetilde{\boldsymbol{l}})\mathrm{d}\widetilde{l},\end{aligned}\tag{A.4}$$

where $\widetilde{\mathbf{N}}:\widetilde{\Omega}\to\mathrm{Sym}$ and $\widetilde{\mathbf{M}}:\widetilde{\Omega}\to\mathrm{Sym}$ are defined by (3.3a) and (3.3b), respectively. In (A.4), $\widetilde{\boldsymbol{l}}$ and $\widetilde{\boldsymbol{t}}$ are unit outward normal and unit (counter-clockwise from $\boldsymbol{e}_3$ axis) tangent vectors of $\partial\widetilde{\Omega}$, respectively, belonging to span$(\boldsymbol{e}_1, \boldsymbol{e}_2)$. These vectors at any $\widetilde{\boldsymbol{x}} \in \partial\widetilde{\Omega}$ are defined by

$$\widetilde{\boldsymbol{l}}(\widetilde{\boldsymbol{x}}) := \frac{\widetilde{\boldsymbol{x}}}{\sqrt{\widetilde{\boldsymbol{x}} \cdot \widetilde{\boldsymbol{x}}}}, \qquad \widetilde{\boldsymbol{t}}(\widetilde{\boldsymbol{x}}) := \boldsymbol{e}_3 \wedge \widetilde{\boldsymbol{l}}(\widetilde{\boldsymbol{x}}). \tag{A.5}$$

Using the relations (1.21), (1.24), the expression (A.4) for $\delta\mathcal{V}_P$ becomes

$$\begin{aligned}\mathfrak{C}^{-1}\delta\mathcal{V}_P = &-\int_\Omega \delta\boldsymbol{v} \cdot (\nabla \cdot \mathbf{N})\mathrm{d}A - \int_\Omega \delta w(\nabla \cdot (\frac{1}{\lambda}\mathbf{N}\nabla w + \nabla \cdot \mathbf{M}))\mathrm{d}A + \\ &+ \int_\Gamma \delta\boldsymbol{v} \cdot (\mathbf{N}\boldsymbol{l})\mathrm{d}l \quad + \int_\Gamma \delta w\big((\frac{1}{\lambda}\mathbf{N}\nabla w + \nabla \cdot \mathbf{M}) \cdot \boldsymbol{l} + \nabla(\boldsymbol{t} \cdot \mathbf{M}\boldsymbol{l}) \cdot \boldsymbol{t}\big)\mathrm{d}l - \\ &- \int_\Gamma (\boldsymbol{l} \cdot \mathbf{M}\boldsymbol{l})(\nabla\delta w \cdot \boldsymbol{l})\mathrm{d}l.\end{aligned} \tag{A.6}$$

In (A.6), $\mathrm{d}A = \mathrm{d}x_1\mathrm{d}x_2$ and $\mathrm{d}l = \mathrm{d}s$ can be chosen for specific calculations. Note that $\nabla$ denotes the differential operator (gradient) with respect to $\boldsymbol{x}$ in the space span$(\boldsymbol{e}_1, \boldsymbol{e}_2)$.

Applying above procedure to $\mathcal{F}_R$, i.e. the potential energy (2.6) of rod along with the terms containing the constraints, the first variation $\delta\mathcal{F}_R$ is found to be

$$\begin{aligned}\delta\mathcal{F}_R = &-\int_{\widetilde{\Gamma}} (-\boldsymbol{n}^L \wedge \widetilde{\boldsymbol{d}_3} + n_{||}\widetilde{\boldsymbol{d}_3})' \cdot \delta\widetilde{\boldsymbol{r}}\ \mathrm{d}\widetilde{l} - \\ &- \int_{\widetilde{\Gamma}} \Big((\boldsymbol{m}_\perp + \boldsymbol{m}_{||})' + (\boldsymbol{n}^L \wedge \widetilde{\boldsymbol{d}_3}) \wedge \widetilde{\boldsymbol{r}}' - \Upsilon\widetilde{\boldsymbol{d}_3} \wedge \boldsymbol{n}^L\Big) \cdot \delta\widetilde{\boldsymbol{\theta}}\ \mathrm{d}\widetilde{l} + \\ &+ \int_{\widetilde{\Gamma}} (\widetilde{\boldsymbol{r}}' \cdot \widetilde{\boldsymbol{d}_3} - 1)\delta n_{||}\mathrm{d}\widetilde{l} + \int_{\widetilde{\Gamma}} (\widetilde{\boldsymbol{r}}' \wedge \widetilde{\boldsymbol{d}_3} + \Upsilon\widetilde{\boldsymbol{d}_3}) \cdot \delta\boldsymbol{n}^L\mathrm{d}\widetilde{l} + \\ &+ \int_{\widetilde{\Gamma}} (\boldsymbol{n}^L \cdot \widetilde{\boldsymbol{d}_3})\delta\Upsilon\ \mathrm{d}\widetilde{l},\end{aligned} \tag{A.7}$$

where we employ the definitions of $\boldsymbol{m}_\perp$ and $\boldsymbol{m}_{||}$ as:

$$\boldsymbol{m}_\perp(\widetilde{s}) \ := \ \beta(\widetilde{\mathrm{R}}(\widetilde{s})\widetilde{\boldsymbol{d}_3}^\dagger(\widetilde{s})) \wedge (\widetilde{\mathrm{R}}'(\widetilde{s})\widetilde{\boldsymbol{d}_3}^\dagger(\widetilde{s})), \tag{A.8}$$

$$\text{and } \boldsymbol{m}_{||}(\widetilde{s}) \ := \ \gamma\widetilde{\boldsymbol{d}_2}^\dagger(\widetilde{s}) \cdot (\widetilde{\mathrm{R}}^\top(\widetilde{s})\widetilde{\mathrm{R}}'(\widetilde{s})\widetilde{\boldsymbol{d}_1}^\dagger(\widetilde{s}))\widetilde{\mathrm{R}}(\widetilde{s})\widetilde{\boldsymbol{d}_3}^\dagger(\widetilde{s}), \tag{A.9}$$

for all $\widetilde{s} \in \widetilde{\mathcal{I}}$, that is (3.5c) and (3.5d), respectively in light of the sentences appearing just before (1.25).

Suppose that all relevant fields with decoration $\widetilde{\ }$ are defined for the rod and identified through (1.27), for example $\boldsymbol{r}(\boldsymbol{x}) = \widetilde{\boldsymbol{r}}(\widetilde{\boldsymbol{x}})$, $\delta\boldsymbol{r}(\boldsymbol{x}) = \delta\widetilde{\boldsymbol{r}}(\widetilde{\boldsymbol{x}})$, $\delta\boldsymbol{\theta}(\boldsymbol{x}) = \delta\widetilde{\boldsymbol{\theta}}(\widetilde{\boldsymbol{x}})$, etc. Recall the fact that $\partial\Omega = \Gamma$ (which also coincides with $\widetilde{\Gamma}$). We choose the reference directions according to the following:

$$\widetilde{\boldsymbol{d}_1}^\dagger(\widetilde{s}) := \boldsymbol{e}_3, \quad \widetilde{\boldsymbol{d}_2}^\dagger(\widetilde{s}) := \boldsymbol{l}(\widetilde{s}), \quad \widetilde{\boldsymbol{d}_3}^\dagger(\widetilde{s}) := \boldsymbol{t}(\widetilde{s}), \tag{A.10}$$

where the unit normal $\boldsymbol{l}$ and the unit tangent $\boldsymbol{t}$ on $\Gamma$ are defined in (A.5). These conditions are equivalent to the assumptions (1.17) for the rod itself.

Using the displacement continuity (1.26), the variation of $\boldsymbol{r}$ is given by

$$\delta\boldsymbol{r} = \delta\boldsymbol{v} + \delta w\boldsymbol{e}_3 \tag{A.11}$$

on $\Gamma$. Substituting (A.11) in (A.7) and simplifying further, it is found that

$$\begin{aligned}\delta\mathcal{F}_R = &- \int_\Gamma \left((\boldsymbol{m}_\perp + \boldsymbol{m}_{||})' + (\boldsymbol{n}^L \wedge \boldsymbol{d}_3) \wedge \boldsymbol{r}' - \Upsilon \boldsymbol{d}_3 \wedge \boldsymbol{n}^L\right) \cdot \delta\boldsymbol{\theta}\, \mathrm{d}l - \\ &- \int_\Gamma \Big(\delta\boldsymbol{v} \cdot ((\mathtt{I} - \boldsymbol{e}_3 \otimes \boldsymbol{e}_3)\boldsymbol{n})' + \delta w(\boldsymbol{n} \cdot \boldsymbol{e}_3)'\Big)\mathrm{d}l + \\ &+ \int_\Gamma (\boldsymbol{r}' \cdot \boldsymbol{d}_3 - 1)\delta n_{||}\mathrm{d}l + \int_\Gamma (\boldsymbol{r}' \wedge \boldsymbol{d}_3 + \Upsilon\boldsymbol{d}_3) \cdot \delta\boldsymbol{n}^L \mathrm{d}l + \\ &+ \int_\Gamma (\boldsymbol{n}^L \cdot \boldsymbol{d}_3)\delta\Upsilon \mathrm{d}l,\end{aligned} \tag{A.12}$$

where $\boldsymbol{n}$ is defined by (3.5a).

The first variation of the whole functional (2.4) is given by the sum of (A.6) and (A.12), i.e. $\delta\mathcal{V} = \delta\mathcal{V}_P + \delta\mathcal{F}_R$. In a manner similar to the statement following (2.4), using the expanded form to emphasize the variables and fields involved, it is clear that the first term on the right hand side is $\delta\mathcal{V}_P \equiv \delta\mathcal{V}_P(\boldsymbol{v}, w; \delta\boldsymbol{v},\ \delta w)$. In a slightly shortened notation, we can also write $\delta\mathcal{V}_P \equiv \delta\mathcal{V}_P[\delta\boldsymbol{v}, \delta w]$. The second term is $\delta\mathcal{F}_R \equiv \delta\mathcal{F}_R(\boldsymbol{v}, w, \mathtt{R}, n_{||}, \boldsymbol{n}^L, \Upsilon; \delta\boldsymbol{v}, \delta w, \delta\boldsymbol{\theta}, \delta n_{||}, \delta\boldsymbol{n}^L, \delta\Upsilon) \equiv \delta\mathcal{F}_R[\delta\boldsymbol{v}, \delta w, \delta\boldsymbol{\theta}, \delta n_{||}, \delta\boldsymbol{n}^L, \delta\Upsilon]$. Thus, $\delta\mathcal{V} \equiv \delta\mathcal{V}(\boldsymbol{v}, w, \mathtt{R}, n_{||}, \boldsymbol{n}^L,\ \Upsilon; \delta\boldsymbol{v}, \delta w, \delta\boldsymbol{\theta}, \delta n_{||}, \delta\boldsymbol{n}^L, \delta\Upsilon)$. In a slightly shortened notation, we can also write $\delta\mathcal{V} \equiv \delta\mathcal{V}[\delta\boldsymbol{v}, \delta w, \delta\boldsymbol{\theta}, \delta n_{||}, \delta\boldsymbol{n}^L, \delta\Upsilon]$.

## Appendix B. Unperturbed equations

We have the following conditions on the boundary which are linear with respect to $\boldsymbol{U}$:

$$\left.\begin{aligned}(\mathtt{I} - \boldsymbol{e}_3 \otimes \boldsymbol{e}_3)\boldsymbol{n}' - \mathbf{N}\boldsymbol{l} = \mathbf{0},\\ \boldsymbol{n}' \cdot \boldsymbol{e}_3 - (\boldsymbol{f} \cdot \boldsymbol{l} + (\nabla \cdot \mathbf{M}) \cdot \boldsymbol{l} + \nabla(M_{r\theta}) \cdot \boldsymbol{t}) = 0,\\ M_{rr} = 0,\end{aligned}\right\} \text{ on } \partial\Omega. \tag{B.1}$$

We have the following equations within the plate:

$$\left.\begin{aligned}
\nabla\cdot\mathbf{N}=\mathbf{0},\\
\nabla\cdot(\boldsymbol{f}+\nabla\cdot\mathbf{M})=0,\\
12\left(\left(\boldsymbol{e}_r\cdot\nabla\boldsymbol{v}\boldsymbol{e}_r+\frac{1}{2}w_{,r}^2\right)+\nu\left(\boldsymbol{e}_\theta\cdot\nabla\boldsymbol{v}\boldsymbol{e}_\theta+\frac{1}{2}w_{,\theta}^2\right)\right)+\\
+6(\frac{1}{\lambda}-1)\left(w_{,r}^2+\nu w_{,\theta}^2\right)+12(1+\nu)(1-\lambda)-\mathfrak{C}^{-1}N_{rr}=0,\\
12\left(\left(\boldsymbol{e}_\theta\cdot\nabla\boldsymbol{v}\boldsymbol{e}_\theta+\frac{1}{2}w_{,\theta}^2\right)+\nu\left(\boldsymbol{e}_r\cdot\nabla\boldsymbol{v}\boldsymbol{e}_r+\frac{1}{2}w_{,r}^2\right)\right)+\\
+6(\frac{1}{\lambda}-1)\left(w_{,\theta}^2+\nu w_{,r}^2\right)+12(1+\nu)(1-\lambda)-\mathfrak{C}^{-1}N_{\theta\theta}=0,\\
6(1-\nu)(\boldsymbol{e}_r\cdot\nabla\boldsymbol{v}\boldsymbol{e}_\theta+\boldsymbol{e}_\theta\cdot\nabla\boldsymbol{v}\boldsymbol{e}_r+w_{,r}w_{,\theta})+\\
+6(1-\nu)(\frac{1}{\lambda}-1)w_{,r}w_{,\theta}-\mathfrak{C}^{-1}N_{r\theta}=0,\\
-\frac{h^2}{\lambda^2}\left(w_{,rr}+\nu\left(\frac{1}{r}w_{,r}+\frac{1}{r^2}w_{,\theta\theta}\right)\right)-\mathfrak{C}^{-1}M_{rr}=0,\\
-\frac{h^2}{\lambda^2}\left(\nu\left(w_{,rr}\right)+\frac{1}{r}w_{,r}+\frac{1}{r^2}w_{,\theta\theta}\right)-\mathfrak{C}^{-1}M_{\theta\theta}=0\\
-\frac{h^2}{\lambda^2}(1-\nu)\left(\frac{1}{r}w_{,r\theta}-\frac{1}{r^2}w_{,\theta}\right)-\mathfrak{C}^{-1}M_{r\theta}=0,\\
\boldsymbol{f}-\frac{1}{\lambda}\mathbf{N}\nabla w=\mathbf{0},
\end{aligned}\right\}\text{ in }\Omega. \tag{B.2}$$

On boundary, we have the following non-linear equations:

$$\left.\begin{aligned}
(\boldsymbol{m}_\perp(\boldsymbol{g},\boldsymbol{g}')+\boldsymbol{m}_{||}(\boldsymbol{g},\boldsymbol{g}'))'-\boldsymbol{n}\wedge\boldsymbol{r}'=\mathbf{0},\\
\boldsymbol{r}'-\boldsymbol{d}_3=\mathbf{0},
\end{aligned}\right\}\text{ on }\partial\Omega. \tag{B.3}$$

## Appendix C. Perturbed equations

On substituting (4.6) in (B.2), the following holds:

$$
\left.\begin{aligned}
\nabla\cdot\mathring{\mathbf{N}}&=\mathbf{0},\\
\nabla\cdot\left(\mathring{\boldsymbol{f}}+\nabla\cdot\mathring{\mathbf{M}}\right)&=0,\\
12\left(\left(\boldsymbol{e}_r\cdot\nabla\mathring{\boldsymbol{v}}\boldsymbol{e}_r+\frac{1}{2}\mathring{z}_{,r}^2\right)+\nu\left(\boldsymbol{e}_\theta\cdot\nabla\mathring{\boldsymbol{v}}\boldsymbol{e}_\theta+\frac{1}{2r^2}\mathring{z}_{,\theta}^2\right)\right)+&\\
+6(\frac{1}{\lambda}-1)\left(\mathring{z}_{,r}^2+\frac{\nu}{r^2}\mathring{z}_{,\theta}^2\right)-\mathfrak{C}^{-1}\mathring{N}_{rr}&=0,\\
12\left(\left(\boldsymbol{e}_\theta\cdot\nabla\mathring{\boldsymbol{v}}\boldsymbol{e}_\theta+\frac{1}{2r^2}\mathring{z}_{,\theta}^2\right)+\nu\left(\boldsymbol{e}_r\cdot\nabla\mathring{\boldsymbol{v}}\boldsymbol{e}_r+\frac{1}{2}\mathring{z}_{,r}^2\right)\right)+&\\
+6(\frac{1}{\lambda}-1)\left(\frac{1}{r^2}\mathring{z}_{,\theta}^2+\nu\mathring{z}_{,r}^2\right)-\mathfrak{C}^{-1}\mathring{N}_{\theta\theta}&=0,\\
6(1-\nu)(\boldsymbol{e}_r\cdot\nabla\mathring{\boldsymbol{v}}\boldsymbol{e}_\theta+\boldsymbol{e}_\theta\cdot\nabla\mathring{\boldsymbol{v}}\boldsymbol{e}_r+\frac{1}{r}\mathring{z}_{,r}z_{,\theta})+&\\
+6(1-\nu)\frac{1-\lambda}{\lambda r}\mathring{z}_{,r}\mathring{z}_{,\theta}-\mathfrak{C}^{-1}\mathring{N}_{r\theta}&=0,\\
-\frac{h^2}{\lambda^2}\left(\mathring{z}_{,rr}+\nu\left(\frac{1}{r}\mathring{z}_{,r}+\frac{1}{r^2}\mathring{z}_{,\theta\theta}\right)\right)-\mathfrak{C}^{-1}\mathring{M}_{rr}&=0,\\
-\frac{h^2}{\lambda^2}\left(\nu\mathring{z}_{,rr}+\frac{1}{r}\mathring{z}_{,r}+\frac{1}{r^2}\mathring{z}_{,\theta\theta}\right)-\mathfrak{C}^{-1}\mathring{M}_{\theta\theta}&=0,\\
-\frac{h^2}{\lambda^2}(1-\nu)\left(\frac{1}{r}\mathring{z}_{,r\theta}-\frac{1}{r^2}\mathring{z}_{,\theta}\right)-\mathfrak{C}^{-1}\mathring{M}_{r\theta}&=0,\\
\mathring{\boldsymbol{f}}-(\frac{1}{\lambda}\mathring{\mathbf{N}}\nabla\mathring{z}+\frac{12}{\lambda}(1+\nu)(1-\lambda)\nabla\mathring{z})&=\mathbf{0},
\end{aligned}\right\}\text{ in }\Omega. \tag{C.1}
$$

Above equations represent the conditions for equilibrium of the plate. Similarly, after substituting (4.6) in (B.1), the boundary conditions are found to be

$$
\left.\begin{aligned}
(\mathtt{I}-\boldsymbol{e}_3\otimes\boldsymbol{e}_3)\mathring{\boldsymbol{n}}'-\mathring{\mathbf{N}}\boldsymbol{l}&=\mathbf{0},\\
\mathring{\boldsymbol{n}}'\cdot\boldsymbol{e}_3-(\mathring{\boldsymbol{f}}\cdot\boldsymbol{l}+(\nabla\cdot\mathring{\mathbf{M}})\cdot\boldsymbol{l}+\nabla(\mathring{M}_{r\theta})\cdot\boldsymbol{t})&=0,\\
\mathring{M}_{rr}&=0,
\end{aligned}\right\}\text{ on }\partial\Omega, \tag{C.2}
$$

while (B.3) leads to

$$
\left.(\boldsymbol{m}_\perp+\boldsymbol{m}_{||})'-(\mathring{\boldsymbol{n}}+6\mathfrak{D}(1+\nu)(\lambda-1)\boldsymbol{t})\wedge\boldsymbol{r}'=\mathbf{0},\boldsymbol{r}'-\boldsymbol{d}_3=\mathbf{0}\right\}\text{ on }\partial\Omega. \tag{C.3}
$$

In (C.3), $\boldsymbol{m}_\perp$, $\boldsymbol{m}_{||}$ and $\boldsymbol{r}$ are given by (3.5c), (3.5d) and (1.26) respectively.

## Appendix D. Simplifications of linearized operator for finding its null space

*Appendix D.1. Domain equations*

The domain equations are simplified to obtain partial differential equations for the determination of $v_r$, $v_\theta$ and $z$. Equating (4.21) to zero and substituting $N_{rr}$, $N_{\theta\theta}$ and $N_{r\theta}$ from $(4.21)_{3,4,5}$ in $(4.21)_1$ and separating components along $\boldsymbol{e}_r$ and $\boldsymbol{e}_\theta$ gives a system of partial differential equations in $v_r$ and $v_\theta$. The system of partial differential equations for $v_r$ and $v_\theta$ is given by

$$\begin{aligned} -2v_r + (1-\nu)v_{r,\theta\theta} + (\nu-3)v_{\theta,\theta} + \\ +r(2v_{r,r} + 2rv_{r,rr} + (1+\nu)v_{\theta,\theta r}) = \ & 0, \end{aligned} \tag{D.1a}$$

$$\begin{aligned} \text{and}\quad (3-\nu)v_{r,\theta} - (1-\nu)v_\theta + \\ +r((1+\nu)v_{r,\theta r} + (1-\nu)v_{\theta,r} + (1-\nu)rv_{\theta,rr}) + 2v_{\theta,\theta\theta} = \ & 0. \end{aligned} \tag{D.1b}$$

Similarly, substituting $M_{rr}$, $M_{r\theta}$, $M_{\theta\theta}$ and $\boldsymbol{f}$ from $(4.21)_{6,7,8,9}$ in $(4.21)_2$ gives a partial differential equation for $z$, given by

$$\begin{aligned} & r\left(r\left(r^2 z_{,rrrr} + 2rz_{,rrr} - z_{,rr} + 2z_{,rr\theta\theta}\right) + z_{,r} - 2z_{,r\theta\theta}\right) + \\ & + 4z_{,\theta\theta} + z_{,\theta\theta\theta\theta} - 12\frac{1+\nu}{h^2}(\lambda-\lambda^2)\left(r^4 z_{,rr} + r^3 z_{,r} + r^2 z_{,\theta\theta}\right) = 0. \end{aligned} \tag{D.2}$$

In (D.2), $v_r$, $v_\theta$ and $z$ are assumed to be sufficiently smooth, such that (D.1) and (D.2) are well-defined.

*Appendix D.2. Boundary conditions*

At the boundary, $r = 1/\alpha$ and $\theta \in \mathcal{L}$. The boundary conditions are given by (C.2) due to the definition of space $\mathcal{D}$ and $(4.21)_{10,11}$ due to the components of the operator $\mathfrak{L}(\lambda)$. Substituting (4.14) and (4.25) in (C.2) and separating polar components gives

$$\alpha n_{r,\theta} - \alpha n_\theta - N_{rr} = 0, \tag{D.3a}$$

$$\alpha n_{\theta,\theta} + \alpha n_r - N_{r\theta} = 0, \tag{D.3b}$$

$$\alpha n_{3,\theta} - (f_r + M_{rr,r} + 2\alpha M_{r\theta,\theta} + \alpha M_{rr} - \alpha M_{\theta\theta}) = 0, \tag{D.3c}$$

$$M_{rr} = 0. \tag{D.3d}$$

Substituting (4.25) in $(4.21)_{10,11}$ and separating polar components gives

$$2\alpha^2\beta s_{r,\theta\theta} - 2\gamma\alpha^2 s_r - 2\alpha^2(\beta+\gamma)s_{\theta,\theta} - 12(1+\nu)(\lambda-1)z_{,\theta} + n_3 = 0, \tag{D.4a}$$

$$\gamma s_{\theta,\theta\theta} - \beta s_\theta + (\beta+\gamma)s_{r,\theta} = 0, \tag{D.4b}$$

$$2\alpha^2\beta s_{3,\theta\theta} + 12(1+\nu)(\lambda-1)(v_{r,\theta} - v_\theta) - n_r = 0, \tag{D.4c}$$

$$\alpha v_{r,\theta} - \alpha v_\theta + 2s_3 = 0, \tag{D.4d}$$

$$v_{\theta,\theta} + v_r = 0, \tag{D.4e}$$

$$\alpha z_{,\theta} - 2s_r = 0. \tag{D.4f}$$

Substituting $s_r$ from (D.4f) in (D.4b) gives an ordinary differential equation in $s_\theta$ as

$$\gamma s_{\theta,\theta\theta} - \beta s_\theta = -\frac{\gamma+\beta}{2}\alpha z_{,\theta\theta}. \tag{D.5}$$

The general solution of (D.5) is given by

$$\begin{aligned} s_\theta(\theta) = & A\exp\left(\sqrt{\frac{\beta}{\gamma}}\theta\right) + B\exp\left(-\sqrt{\frac{\beta}{\gamma}}\theta\right) - \\ & - \alpha\frac{\gamma+\beta}{2}\left((\gamma\tfrac{d^2}{d\theta^2} - \beta)^{-1} z_{,\theta\theta}(\frac{1}{\alpha},\theta)\right). \end{aligned} \tag{D.6}$$

Assuming $z \in \mathcal{C}^2_{\mathcal{S}}$ and applying periodicity condition over $\mathcal{L}$, the first two terms in (D.6) are zero, i.e., $A = B = 0$.

Substituting $M_{rr}$, $M_{r\theta}$, $M_{\theta\theta}$ and $\boldsymbol{f}$ from $(4.21)_{6,7,8,9}$, $n_3$ from (D.4a), $s_r$ from (D.4f) and $s_\theta$ from (D.6) in (D.3c) gives one of the boundary condition for $z$.

$$\begin{aligned}&2\alpha^3\beta s_{r,\theta\theta\theta} - 2\gamma\alpha^3 s_{r,\theta} - 2\alpha^3(\beta+\gamma)s_{\theta,\theta\theta} - \alpha 12(1+\nu)(\lambda-1)z_{,\theta\theta} + \\ &+ (f_r + M_{rr,r} + 2\alpha M_{r\theta,\theta} + \alpha M_{rr} - \alpha M_{\theta\theta}) = 0.\end{aligned} \tag{D.7}$$

The second boundary condition for $z$ is obtained by substituting $M_{rr}$ from $(4.21)_6$ in (D.3d).

Substituting $s_3$ from (D.4d) in (D.4c) and simplifying gives $n_r$ in terms of functions $v_r$ and $v_\theta$ as

$$n_r = \alpha^3\beta(v_{\theta,\theta\theta} - v_{r,\theta\theta\theta}) + 12(1+\nu)(\lambda-1)(v_{r,\theta} - v_\theta). \tag{D.8}$$

Differentiating (D.3a) with respect to $\theta$ and adding (D.3b) and substituting $N_{rr}$, $N_{r\theta}$ and $n_r$ from $(4.21)_{3,5}$ and (D.8), respectively gives one of the boundary condition for $v_r$ and $v_\theta$. It is given by

$$\alpha n_{r,\theta\theta} + \alpha n_r - N_{rr.\theta} - N_{r\theta} = 0. \tag{D.9}$$

The second boundary condition for $v_r$ and $v_\theta$ is given by (D.4e).